\documentclass[aps,prd,twocolumn,groupedaddress,showpacs,nofootinbib,amssymb]{revtex4}

\usepackage[T1]{fontenc}
\usepackage[latin1]{inputenc}
\usepackage{graphicx}
\usepackage[english]{babel}
\usepackage{graphicx}
\usepackage{epstopdf}
\usepackage{bm}
\usepackage{amsmath}
\usepackage{amssymb}
\usepackage{amsfonts}
\usepackage{epsfig}
\usepackage{colordvi}
\usepackage{color}
\usepackage{amsthm}

\renewcommand*{\Re}[1]{\operatorname{Re}\left\{#1\right\}}

\begin{document}

\title{Majorana quanta, string scattering, curved spacetimes and the
Riemann Hypothesis}

\author{Fabrizio Tamburini} 
\email{fabrizio.tamburini@gmail.com}
\affiliation{ZKM -- Zentrum f\"ur Kunst und Medientechnologie, Lorentzstr. 19, D-76135, Karlsruhe, Germany.}

\author{Ignazio Licata}
\email{ignaziolicata3@gmail.com}
\affiliation{Institute for Scientific Methodology (ISEM) Palermo Italy}
\affiliation{School of Advanced International Studies on Theoretical and Nonlinear Methodologies of Physics, Bari, I-70124, Italy}
\affiliation{International Institute for Applicable Mathematics and Information Sciences (IIAMIS), B.M. Birla Science Centre, Adarsh Nagar, Hyderabad -- 500 463, India}

\begin{abstract} 
The Riemann Hypothesis states that the Riemann zeta function $\zeta(z)$ admits a set of  ``non-trivial'' zeros that are complex numbers supposed to have real part $1/2$. Their distribution on the complex plane is thought to be the key to determine the number of prime numbers before a given number. 
Hilbert and P\'olya suggested that the Riemann Hypothesis could be solved 
through the mathematical tools of physics, finding a suitable Hermitian or unitary operator that describe classical or quantum systems, whose eigenvalues distribute like the zeros of $\zeta(z)$. 
A different approach is that of finding a correspondence between the distribution of the $\zeta(z)$ zeros and the poles of the scattering matrix $S$ of a physical system. 
Our contribution is articulated in two parts: in the first we apply the infinite-components Majorana equation in a Rindler spacetime and compare the results with those obtained with a Dirac particle following the Hilbert-P\'olya approach showing that the Majorana solution has a behavior similar to that of massless Dirac particles and finding a relationship between the zeros of zeta end the energy states.
Then, we focus on the $S$-matrix approach describing the bosonic open string scattering for tachyonic states with the Majorana equation.
Here we find that, thanks to the relationship between the angular momentum and energy/mass eigenvalues of the Majorana solution, one can explain the still unclear point for which the poles and zeros of the $S$-matrix of an ideal system that can satisfy the Riemann Hypothesis, exist always in pairs and are related via complex conjugation. 
As claimed in the literature, if this occurs and the claim is correct, then the Riemann Hypothesis could be in principle satisfied, tracing a route to a proof.
\end {abstract}

\pacs{02, 0.3.65.Pm, 04.20.-q, 04.50.-h, 11.25.-w, 11.25.Wx}
%
%
%
\maketitle
%
%

\section{Introduction} 
The Riemann Hypothesis (RH) and the distribution of prime numbers can be considered at all effects the Holy Grail of Number Theory and Mathematics. 
Bernhard Riemann in 1859 hypothesized that the distribution of the non-trivial zeros of the function $\zeta(z)$, of the complex variable $z \in \mathbb{C}$, the analytic continuation of the sum of Dirichlet's series
\begin{equation}
\zeta(z)= \sum_{n=1}^{+ \infty} \frac1{n^z},
\end{equation}
would be the key to describe the distribution of the prime numbers before a given number \cite{riemann1,edwards,gram,hutch}. 
As defined by Riemann, the zeros of $\zeta(z)$ are the correction terms used to determine, through the prime-counting function $\pi(x) \sim Li(x)$, where the function $Li(x) = \int_2^x \frac {dt}{\ln(t)}$ is the logarithmic integral function, 
how many primes there are up to - and including - any given number $x$. 
For $x=1$ then $\pi(1)=0$, for $x=2$ one obtains $\pi(2)=1$ and for $x>3$ one can use, e.g., the formula by Hardy and Wright, derived from the Waring - Wilson - Lagrange  theorem  \cite{HW}, for which the number of primes before the given number $n$ can be given by
\begin{equation}
\pi(x)= -1 + \sum^x_{k=3} \left\{ (k-2)! - k~\mathrm{floor} \left[ \frac{(k-2)!}{k} \right] \right\}
\end{equation}
where ``floor~(x)'', also written as $\lfloor x \rfloor = \mathrm{sup}\{m \in \mathbb{Z}~|~ m \leq x \}$, is the floor function that gives the largest integer less than or equal to a given number \cite{floor} and ``sup'' is the supremum, the least upper bound of the interval $(3,x]$ that must be integer $(\mathrm{i.e.,}~m\in \mathbb{Z})$.

The auxiliary function $J(x)$ of the prime counting function $\pi(x)$, defined by Riemann, is related with $\pi(x)$ through the M\"obius function $\mu(k)$ \cite{edwards},
\begin{equation}
\pi(x)=\sum_{k=1}^{\infty} \frac{\mu(k)}{k} J(x^{1/k})
\end{equation}
and Riemann proposed for $J(x)$ the formulation reported below, where the role of the zeros of the zeta function in the counting of primes is more evident
\begin{equation}
J(x) = Li(x) - \sum_\rho Li(x^\rho) - \log 2 +\int_2^\infty \frac {dt}{t( t^2 -1) \log t}
\label{counting}
\end{equation}
here the $\rho$'s are the nontrivial zeros of the $\zeta(z)$ function. 
The sum of powers of $x$ with exponents the $\zeta(z)$'s zeros $\rho$, in the second term of Eq.~(\ref{counting}), is a correction to the overestimation of the number of primes given by the first term, the logarithmic integral function, and presents a peculiar oscillating behavior that is also known as ``the music of primes''. 
This is evident in the physics approach to the Riemann hypothesis, where the $\zeta$ function is found in different branches of physics, from classical to quantum mechanics, statistical physics, quantum field and string theory, where the Riemann zeta is a key function to describe certain physical models \cite{wolf}. An example is found in the Hilbert-P\'olya approach where were studied classical chaotic and quantum dynamical systems with Hamiltonians based on the product of the position $x$ and momentum $p$. The music of primes is there related to the fluctuations about the smooth mean density of zeros that reflect the classical chaos, associated with the distribution of primes \cite{m1,m2,m3,schumi}.

While the trivial zeros of the function $\zeta(z)$ are negative even integers that represent a countably infinite (denumerable) set of numbers isomorphic to that of natural numbers, $\mathbb{N}$, the non-trivial zeros of $\zeta(z)$ form instead an infinite countably set of complex numbers $z \in \mathbb{C}$, whose real part is supposed to be always $\Re z= 1/2$. Thus, the non-trivial zeros of $\zeta(z)$ are expected to lay in the so-called ``critical line'' $\sigma = 1/2$ of the complex plane. The Riemann Hypothesis (RH) is valid if all nontrivial zeros of $\zeta(z)$ are found on the critical line. Hardy showed in 1914 that the zeros on the critical line form an infinite and denumerable set of complex numbers, without excluding a-priori the existence of at least one nontrivial zero outside the critical line that could invalidate the Riemann Hypothesis \cite{hardy}. 

To prove the RH, one can use the mathematical tools and concepts utilized in Physics and adopt the approach suggested by Hilbert and P\'olya around the years 1912--1914:
``\textit{The Riemann Hypothesis is true if there exists a positive operator}\footnote{Given a Hilbert space $\mathbb{H}$ and an operator A $\in$ L($\mathbb{H}$), A is said to be a positive operator if $\langle \mathrm{A} x, x\rangle \geq 0,~\forall x \in$  $\mathbb{H}$. A positive operator on a complex Hilbert space is a symmetric operator with self-adjoint extension that is also a positive operator and are Hermitian operators.}  \textit{whose spectrum of eigenvalues finds a $1-1$ correspondence with all the elements of the set of the non-trivial zeros of} $\zeta(z)$'', no matter what  the physical system this operator is describing. To be more precise, the operator can be also complex-Hermitian or real, as the zeros of $\zeta$ can assume both positive and negative imaginary values.
On this line, Montgomery showed that, assuming the validity of the Riemann hypothesis, the imaginary part of the Riemann zeros would satisfy the statistics of a Gaussian unitary ensemble (GUE) \cite{montgomery}. Then, in 1987 Odlyzo computed a large number of zeros and found a deviation of the GUE law \cite{odly} that were explained by Berry and collaborators with quantum chaos theory \cite{berrych1,berrych2}.
Thus, if there exists a linear operator $\hat A$, whose eigenvalues $a_k$ distribute accordingly to that of the zeros of $\zeta(z)$, this linear operator is expected to be complex Hermitian or unitary as occurs for the eigenvalues of a random complex Hermitian operator with eigenvalues $1/2 + i \lambda_n$ or with those of a large order unitary matrix that have precisely the same pair correlation function with, instead, real eigenvalues.

Alternatively, the Riemann hypothesis is valid also if \textit{the set of nontrivial zeros of the associated Riemann function} 
\begin{equation}
\xi(z)= \frac 12 z (z-1) \pi^{-z/2} \Gamma \left( \frac z2 \right) \zeta(z)
\end{equation}
\textit{correspond to the eigenvalues of a positive operator} of a given physical system.
The function $\xi(z)$ is an entire $n-$differentiable ($n \leq \infty$) complex-valued and holomorphic function defined in the complex field $\mathbb{C}$, i.e., $\xi(z) \in C^{n\leq\infty}_\mathbb{C}$, and real-valued when the input $z$ is real, viz., $z\in\mathbb{R} \Rightarrow \xi(z) \in\mathbb{R}$.

Berry and Keating \cite{berry1,berry2} and Connes \cite{connes} initially proposed the spectral realization of the Riemann zeros through the quantization of the classical Hamiltonian $H = xp$ of a system with positive-defined energy $E$ then extended to quantum systems \cite{schumi}.
The quantities $x$ and $p$ represent the position and momentum of a classical particle moving in the space described by an unidimensional spatial coordinate $x$ of a spacetime $(t,x)$ with dimensions $D=1+1$. The quantity $xp$ can be also interpreted in terms of angular momentum projected in $(t,x)$. 

The first steps made to build a connection between the eigenvalues of the Hamiltonian $H = xp$ and the spectrum of Riemann's $\zeta(z)$ zeros were based on the regularization schemes obtained by dividing -- or better, ``sawing'' -- in sub-regions the phase space by introducing a geometric texture made of rectangles with sides $l_x$ and $l_p$ and area $l_x~l_p = 2\pi \hbar$, such that $|x| > l_x$ and $|p| > l_p$. This structure has properties similar to that of a Planck cell in the phase space that recalls Heisenberg's uncertainty principle.

The classical trajectories in the phase space of this type of chaotic dynamical system, given the initial conditions $x_0$ and $p_0$, can then be written as follows
\begin{equation}
x(t) = x_0e^t ; \,\,\, p(t) = p_0 e^{-t}; \,\,\, E = x_0p_0,
\end{equation}
and the energy $E$ draws a hyperbola in the phase space.
At a first sight, one can find that the T--symmetry is broken and it could result a problem for the demonstration of the RH if, by inverting $t \rightarrow - t$, one finds $x \rightarrow x$ and $p \rightarrow -p$. 
This is not the case, since it is not the standard formulation of the nonrelativistic energy, for which $p = \dot x$. In fact, this is a constrained system without an explicit (or any) relation between velocity $\dot x$ and momentum $p$.
The negative energy can be found in several examples such as the classical inverted oscillator, which is known to present a chaotic behavior. In this specific case it is related with $xp$ through a rotation in the phase plane, a classical system that does not require the formulation of a Dirac negative energy sea of fermions.
In this case, these quantities draw trajectories characterized by negative values of the energy, $E \rightarrow - E$; this is a physical condition that can 
be found in quantum mechanics. A known example of negative energy states that can be found in quantum mechanics is the sea of negative states of electrons discussed in the early interpretations of the energy spectrum of Dirac's equation \cite{dirac}.

Anyway, to deal with negative energy states is not a problem in view of a verification of the RH with the Hilbert-P\'olya approach. 
An example is the black hole $S$-matrix by  't Hooft, which can be either expressed in terms of an auxiliary problem of inverted Harmonic Oscillator scattering or as the Dilation Operator close to the BH horizon \cite{betzios}. This approach to the RH is at all effects related to the Hilbert P\'olya conjecture and the approach made by Berry and Keating to the RH \cite{betzios2}. 
The needed appropriate phase space conditions that discretize the spectrum of the operator are given by an appropriate form of gauging of charge, parity and time (CPT). The non-trivial zeros then correspond to the real discrete spectrum of the quantum Hamiltonian, whilst the trivial zeros correspond to unstable resonances.

Switching to quantum mechanics, one can in principle obtain Hermitian Hamiltonians from the classical $H=xp$ by symmetrizing $(+)$ the product of the operators $\hat x \hat p$,  obtaining $\hat H_+ = k/2 \left( \hat x \hat p + \hat p \hat x \right)$, where $k$ is a real arbitrary number and, including the anti symmetrized term $(-)$, one obtains $\hat H_\pm = k/2 \left( \hat x \hat p \pm \hat p \hat x \right)$ that can be summarized in a more elegant way through a phase factor $e^{i \alpha}$ with $\alpha \in \mathbb{R}$. As $\hat H$ is invariant with respect to $\lambda \in \mathbb{R}$ thanks to the Poincar\'e transformation, $x \rightarrow \lambda x$ and $p \rightarrow \lambda^{-1} p$; for the sake of simplicity one can set $k=1$ and write, without loss of generality,

\begin{equation}
\hat H_\alpha = \frac 12 \left( \hat x \hat p + e^{i \alpha} \hat p \hat x \right).
\end{equation}
For $e^{i \alpha}= 1$ one obtains the symmetric Hamiltonian used by Berry and Keating for the RH,
\begin{equation}
\hat H_+ = \frac 12 \left( \hat x \hat p + \hat p \hat x \right) = -i ~\left( \hat x \frac d{dx} + \frac 12  \right).
\end{equation}
when, instead, $e^{i \alpha}= -1$, the Hamiltonian becomes the antisymmetric one,  $\hat H_- =  \frac 12 \left( \hat x \hat p - \hat p \hat x \right)$, which is not Hermitian and leads to the class of PT-symmetric quantum Hamiltonians that can have either complex conjugate or purely real eigenvalues. 
In general, when the Hamiltonian is not Hermitian and presents PT-symmetry, one can argue that in general $\hat H$ may admit an infinite number of positive real eigenvalues but it cannot be a-priori excluded the existence of a single complex value that falsifies the approach to the RH. Exceptions are made when one finds a specific PT-symmetric system with real eigenvalues \cite{devincenzo,El-Ganainy,Mannheim}. 
Another issue that one can find is that certain classes of quantum Hamiltonians can have the inner-product space of eigenstates that is not a usual Hilbert space and the elements of the vector space may have infinite norm. 

This problem recalls what happens in quantum field theories, the infinite norm of state vectors are the result of the use of an infinite and in general flat spacetime volume. Formulated differently, that type of norm can be interpreted as a sort of infrared divergence. 
One way to face this problem is to introduce a finite interval of spacetime and use an integral representation of the delta function that characterizes the norm. 
The terms 
that may appear in the norm can be replaced by the volume interval, cancelling the divergences and adopting a hyperbolic geometric support for which one imposes a log-behavior like in a Rindler spacetime. Some interesting remarks can be found in Ref. \cite{strumia} with PT-symmetric quantum mechanics. 

Interestingly, all this does not apply when the Hamiltonian is formulated in a way that all eigenvalues are indeed real and correspond to the nontrivial zeros on the critical line. That is, the operator is indeed self-adjoint, and there cannot be isolated exceptions. 
This is obtained by facing the issue of normalization with a Rigged Hilbert space formalism \cite{antoine} as can be found in the approach of Ref. \cite{bender} and briefly discussed in the Appendix (\ref{appA}).
Another possible way to approach this problem could be to use the so-called 4-derivative oscillator $q(t)$, a tool used in quantum gravity where the proposal of a renormalizable extension of general relativity can be formulated adding to the Einstein-Hilbert Lagrangian some quadratic terms in the curvature tensor. In this view, the graviton has a 4-derivative term, as $q(t)$ mimics massless graviton and massive graviton ghost modes with linear coupling with tensor matter fields. This theoretical construct has positive energy eigenvalues, normalizable wavefunctions, unitary evolution in a negative-norm configuration space.

Here, the 4-derivative oscillator $q(t)$ leads to a formalism that recalls the deterministic part of quantum mechanics. This can be summarized as follows: positive energy eigenvalues, normalizable wave-functions and a time evolution operator, $\hat U = e^{-i \hat H t}$, which preserves the indefinite quantum norm if the Hamiltonian $\hat H$ is self-adjoint \cite{salvio}. 
What is also interesting is that in the 4-derivative oscillator, when decomposed in the two modes system, one mode has a positive classical Hamiltonian and the other one, instead, corresponding to the ``ghost'' mode, has negative classical Hamiltonian. 
If we consider the simplest case of an Indefinite-Norm Free Harmonic Oscillator, the quantization of the ghost can be made with the Dirac-Pauli representation and can be made positive in the Majorana formulation after some transformations.

\subsection*{Summary of the present work}
In this work, after this introductory review on the Hilbert and P\'olya approach to the Riemann Hypothesis, in section II we present the results obtained by using the infinite-components Majorana equation instead of the Dirac equation in a Rindler spacetime describing uniformly accelerated quanta as in Ref. \cite{sierra}, finding that the massive Majorana particles, either bosonic or fermionic, with always a positive-defined mass, behave like massless Dirac fermions. 
Moreover, the equations for the eigenenergies $E_n$ that describe the Majorana particles show a relationship, through the Mellin-Barnes integral representation of the modified Bessel function of the second type, $K_\nu(x)$, with the Riemann's zeta function, leading to a relationship between the zeros of $K_\nu(x)$ and those of $\zeta(x)$.

In section III we focus, instead, on the scattering $S$-matrix approach to the Riemann Hypothesis applying the Majorana conditions to the formalism of bosonic string scattering.
We find that, by assuming the correctness of the claim in Ref. \cite{castro}, where to demonstrate the validity of the Riemann Hypothesis one needs to explain the reason why the zeros of the complex double poles of the $S$-matrix of a physical system always appear in complex-conjugate pairs, the requirements to prove the RH can be in principle considered satisfied.

In other words, we find that if the initial assumptions are valid, the string scattering of tachyonic quanta and strings obeying the infinite-components Majorana equation can 
provide this explanation and satisfying this requirement, it draws a route to the proof for the Riemann Hypothesis.

Section III ends with a short analysis on the $S$-matrix pole dynamics, followed by the conclusions and the appendix where PT-symmetric systems are briefly reviewed together with the Majorana infinite-components equation, including additional suggestions for a deeper insight in the bibliography. The work is ended with a brief discussion on Hardy's Z function and Wick's rotations.

\section{$H=xp$ and The Majorana Tower}
The Hilbert-P\'olya approach to the RH consists in the search or, better, in the formulation of a suitable operator like the Hermitian Hamiltonian of a physical system, whose eigenvalues find a correspondence with the distribution of the Riemann's zeta zeros.

Our contribution in this section is to apply the Majorana infinite-components equation instead of the Dirac equation in the Hamiltonian of a quantum system and find that the equivalence between particle and antiparticle together with the positive-defined condition of the energy of the Majorana particles draws a behavior similar to that of a massless Dirac particle.

The class of $H=xp$ Hamiltonians, based on the $xp$ product, present a distribution of eigenvalues that can mimic that of the Riemann zeros efficiently. 
At our present knowledge, Hermitian Hamiltonians of the class $H=xp$ and their variants considered in the literature give interesting approaches to the demonstration of the Riemann Hypothesis but, up to now, there has not been a clear trace of the exact distribution of the Riemann zeros in the spectra of the $H=xp$ Hamiltonians.
In general, one finds that either the mean eigenvalue density differs from that of $\zeta(z)$ zeros and present fundamental differences in the periodic orbits \cite{berry11}, or because the Hamiltonians are not Hermitian as is thought to be strictly required by the Hilbert-P\'olya approach.
Only certain particular examples seem to be very close to satisfy the Hilbert-P\'olya approach even if there is still a lot of exciting research to be done \cite{bender,sierra0}.

We now focus on the analysis of the motion of uniformly accelerated quanta. 
In the search of an operator able to satisfy the Hilbert-P\'olya approach, a further improvement to mimic efficiently the $Li(z)$-type behavior of $\pi(x)$ and of the auxiliary function $J(x)$ in Eq.~(\ref{counting}), is made by adopting a system that describes a uniformly accelerated particle with Hamiltonian
\begin{equation}
H_R = x \left( p + \frac vp \right),
\label{hsierra}
\end{equation}
as its eigenvalues present a log-type distribution; $x$ and $p$ are the position and momentum of the particle and $v = m^2 \in \mathbb{R}$ represents the squared mass of the particle.

Using a more precise formal language, this improvement is obtained by modifying the geometry of space and time with General Relativity: the uniformly accelerated particle is described by the Dirac equation in a Rindler spacetime with $D = 1+1$ dimensions \cite{sierra} and this study has been extended to a spacetime with $D = 3 + 1$ dimensions \cite{maa}. 
The Rindler spacetime is a hyperbolically accelerated reference frame that is comparable to a homogeneous gravitational field, a solution of Einstein's equations in General Relativity, used to study several fundamental physical problems like the description of a free falling radiating charge \cite{rindler,freefall}. 
This manifold acts as a Minkowski spacetime with coordinates adapted to a boost-Killing vector field and is equivalent to a uniformly accelerated observer where all Lorentz boosts, rotations and their combinations are continuously connected to the identity operator.

Interestingly, one can generalize $H_R$ with a general function $v(q)$ defined in the field of real or complex numbers $q \in \mathbb{C}$  instead of the constant parameter $v$, to write a more general locally hyperbolic metric described through Lorentz boosts by a family of Rindler spacetimes or a family of Rindler observers $R_q$, that can better overlap the eigenvalues of $H_R$ with the distribution of the zeros of $\zeta(z)$ or, equivalently, with the use of a function $V(x)$ of the spatial coordinate acting as a potential, as discussed in the literature, giving 
\begin{equation}
\hat{H}_R = U(x) p - \frac vp V(x). 
\end{equation}
Physically this can be in principle achieved also by varying a property that characterizes the moving particle, like the mass, angular momentum and so on, which results equivalent to the introduction of the effect of a potential in its dynamics.

The Dirac theory of a relativistic particle in a Rindler spacetime provides the relativistic quantum correspondent solution of the modified $H=xp$ - class Hamiltionians: the spectrum of a Dirac massive fermion in the domain of a D=1+1 Rindler spacetime for which the spatial coordinate is larger or equal to the energy, agrees with the average Riemann zeros. A clear description can be found in the works by Sierra \cite{sierra,sierra2}.

The spectrum of energy eigenvalues of the quantum Hamiltonian $\hat H_R$ obtained from Eq.~(\ref{hsierra}) was used to describe the physics of Dirac fermions in a $D = 1 + 1$ dimensional Rindler spacetime. This spectrum actually obtains a smooth approximation of the Riemann non-trivial zeros for massive Dirac particles, as it finds better agreement with the distribution of the zeros of the ``fake'' P\'olya zeta function. 
Interestingly, this model finds a better fit with the zeros of zeta when one considers accelerated zero-mass fermions like Majorana neutrinos, i.e., when particles are their own antiparticles.

We now apply the infinite-components Majorana solution to the description of uniformly accelerate quanta in a Rindler spacetime. Of course, in this approach, second-quantization and vacuum effects like the Unruh effect are neglected: a Rindler observer would experience the vacuum as a bath of real pairs of entangled particle-antiparticle, which is perceived as thermal radiation \cite{rosabal}.

The Majorana formulation presents a clear relationship between the dynamics of the particles and intrinsic properties of these quanta, like the spin angular momentum $j$ and the mass $m$ and energy $E$. 
The set of infinite-components that build up  the Majorana solution -- also known as the ``Majorana Tower'' \cite{Majorana:NC:1932,magueijo} -- is characterized to have the energy and mass of the particle that depend on the particle's angular momentum $j$ that does not have mandatorily discrete values \cite{beka,suda} (and vice-versa). 
A short introduction on the Majorana Tower can be found in Appendix \ref{appB}.
For a brief discussion on the properties of Dirac and Majorana particles in Rindler coordinates see Ref. \cite{rohim}.

Noteworthy, in the Majorana solution fermions and bosons are treated in a completely equivalent way and are charge-conjugation invariant. This occurs also for single particle states and superpositions of particles with real and imaginary positive and negative values of the mass \cite{nanni}.

We now introduce the properties of Majorana particles starting from the Dirac equation.
If one considers a Majorana particle in the Majorana representation the Dirac operator, the complex operator acting on the wavefunction, becomes a real operator and gives real-valued solutions to the Dirac equation.
It is well known from the standard literature that also a particle described by the real-valued solution of the Dirac equation is a Majorana particle and it has to be a massive neutral fermion or boson that coincides with its own antiparticle.
As an example, consider the equation that describes such a particle for the spin value $s=1/2$ is given by the standard Dirac equation 
\begin{equation}
i \hbar \hat{1}_2 \partial_t = \hat h \Psi \, ,
\label{direq}
\end{equation}
and $\hat{1}_2$ is the $2 \times 2$ identity operator; $\hat h$ is the Hamiltonian which is unbounded and Hermitian but cannot act on all the wave functions of the Hilbert space of the system \cite{devincenzo} unless resizing the Hilbert space as in Ref. \cite{bender}.
Differently from the Dirac equation in Eq.~(\ref{direq}), the wavefunction for a Majorana particle has to be real-valued, $\Psi = \Psi^*$, and has to obey the charge-conjugation condition, $\Psi_C$. 
The formulation that extends this approach to other spin values is obtained in the generalization of the Dirac equation written by Majorana in 1932 \cite{Majorana:NC:1932}, also known as the Majorana Tower is based on infinitesimal Lorentz transformations obtained through a non-unitary transformation and Hermitian operators that generate space-space and time-space relations with infinite components. This is indeed possible when the wave function transforms under unitary representations of the homogeneous Lorentz group and these representations are infinite dimensional \cite{casalbuoni}.

Let us briefly introduce the main properties of the Majorana infinite-components equation that can be of interest to us now. By definition, the energy of a particle at rest in the lab reference frame is $E_0=mc^2$. In the Majorana Tower, particle and antiparticle states coincide and the energy $E$ results also connected with the spin angular momentum $j$ of the particle, giving rise to an infinite spectrum of particles with different energy and spin values,
\begin{equation}
E=\frac{m c^2}{j+ \frac 12},
\label{spinmass}
\end{equation}
with the result of generating a spectrum of different mass values too, also known as the Majorana mass term
\begin{equation}
M^*_j=\frac{m}{j+ \frac 12}.
\label{majmass}
\end{equation}

Considering our case, when we analyze a single particle in a $D = 1 + 1$ spacetime, $\Psi_C$ describes the particle's state with negative energy even if the Majorana condition continues to provide a real solution for the time-dependent Dirac equation.
Regarding the requirements to satisfy the RH, any wave equation of first order in the time derivative can have real solutions implying a  purely imaginary Hamiltonian, which is formally self-adjoint and its mean value -- in a real-valued state -- vanishes.
This is valid also for the momentum operator and when the Hamiltonian is unbounded, one can describe the evolution of the system through an evolution operator as usual.

The example of a Majorana particle in a box can show that the Majorana quantum field operator is Hermitian because the wave function $\Psi$ of the field is real. 
In a Rindler spacetime $x^0=t =\rho \sinh \phi$ and $x^1 = x =  \rho \cosh \phi$, the general Majorana solution with mass $m$ and spin $j$ (assuming $\hbar = c = 1$) of a particle initially at rest in a box with size $L$ has the form
\begin{equation}
\begin{split}
\Psi_\mp (\rho, \phi) = K \sqrt{\frac 1L} \sum_{2j=1}^{+ \infty} \sum_{p=- \infty}^{+ \infty} \sqrt{\frac m{E_p }} a_j \left\{ b_p \psi(p,s) \right. \\
e^{[ - \rho \cosh (\beta-\phi)m/(j+1/2)]} \left.  + c.c.\right\}
\end{split}
\label{majbox}
\end{equation}
\\
and $K$ is a normalization constant, $\sum_{p= - \infty}^{+ \infty} | b_p |^2 = 1/2$ and 
$\sum_{2j=1}^{+ \infty} |a_j|^2 = 1$.

As a Majorana particle with a fixed spin value $j$ can be written in terms of a superposition of a particle and its antiparticle, particle and antiparticle states must coincide and $\hat \Psi = \hat \Psi^\dagger$ also in particular cases like those described in Eq.~(\ref{majbox}).
From this, we can obtain a relationship of the energy eigenvalues whose distribution can fit with that of the zeros of Riemann's zeta obtained from the continuous superposition of plane wavefunctions as discussed in Ref. \cite{sierra} for Dirac particles where, interestingly, for Dirac particles, a correspondence with the fake P\'olya zeta function $\xi^*$ and the Riemann $\xi$ function, described in Ref. \cite{edwards}, has been already found.


When applied to a Majorana particle in a $D = 1 + 1$ Rindler spacetime, as the energy $E$ must be positive-defined as required by the Majorana solution, one finds that chirality has to be taken in account too. In these $1 - D$ spatial structures, as in our case, chirality means mirror reflection and rotation is not defined.

In $D = 1+1$ spacetime, for a standard Dirac particle, antimatter is the mirror-reflected, or chiral, correspondent particle: antimatter is matter evolving its dynamics backwards in time. The two solutions to the Dirac equation are two particle states propagating in the positive and negative x- directions (left and right fermion modes) and related together, through the non zero mass term.

The infinite-components Majorana solution \cite{Majorana:NC:1932} remains valid also in $D=1+1$ as it is based on the Lorentz group of infinitesimal transformations with the additional constraint that the energy of particle and antiparticle states must coincide, as any particle coincides with its antiparticle. This latter condition imposes a very different behavior to Majorana particles with respect to Dirac ones.
Because of this, the Majorana solution together with the solutions away from the Weil equations that are valid for massless particles, behave as the were two different particles \cite{zra}:
\begin{eqnarray}
i \left( \hat{\sigma}^\mu \partial_\mu \right) \Psi_R - M \epsilon \Psi_R^* = 0 \nonumber
\\
i \left( \sigma^\mu \partial_\mu \right) \Psi_L + M' \epsilon \Psi_L^* = 0
\label{majarray}
\end{eqnarray}
here $\hat{\sigma}^\mu$ is the correspondent complex-conjugated matrix of the Pauli matrix $\sigma^\mu$ and $\Psi_R$ and $\Psi_L$ are the positive and negative helicities related to the propagation along the positive and negative spatial axis.
The two equations in the system of equations Eq.~(\ref{majarray}) actually describe two completely different objects with Majorana masses $M$ and $M'$ that do not possess any additive quantum numbers and particles are their own antiparticles but for a phase factor $\theta$, as expected from particles like Majorana neutrinos. The matrix $\epsilon$ is defined as
$\epsilon = \left(\begin{array}{cc} 0 & 1 \\ -1 & 0 \\ \end{array}\right)$.
As any Majorana particle coincides with its antiparticle, one obtains the following identity, a condition derived from the energy eigenvalues of the solutions to Eq.~(\ref{majarray}) for the positive-defined value of the energy $E$ and the other independent solution with energy $-E$, for which the negative sign in the energy indicates a different state of chirality with respect to the positive solution.
In the case of of a bradyonic (v<c) particle or a particle in the lab reference frame from the boundary conditions from the energy eigenvalues for each of the two different Majorana particles, with the energy values $E$ labeled by $(+)$ and $(-)$ for the $L$ and $R$ chirality states, one finds
\begin{equation}
\left(e^{i\theta}- 1 \right) K_{\frac 12 \pm \frac{i E}2}\left(\frac m{a\left(j+ \frac 12\right)} \right) = 0 
\label{majoranarindler}
\end{equation}
and $E= E_0/(j+1/2)$, with $E_0=m$ is the energy in the lab reference frame of a Rindler particle with acceleration $a$ when $c=1$. 
Here one obtains a result similar to that of two independent zero-mass Dirac particles oppositely propagating in a Rindler spacetime when $j=1/2$, but for a phase term $\theta$~ \cite{sierra} this is at all effects the equation for a Majorana neutrino and represents a way to tell the difference between a Fermi standard neutrino and the Majorana one \cite{zra}.
We recall that the left $L$ and right $R$ - propagating solutions, labeled with $(+)$ and $(-)$, are two different particle states and each of these particles that belong to the the Majorana Tower can be written in terms of a superposition of particle and antiparticle states for each $L$ or $R$ independent solutions, as particle and antiparticle states coincide because of the Majorana condition \cite{nanni}.

The identity in Eq.~(\ref{majoranarindler}) remains valid also when are superimposed all the particle states of the whole Tower of particles obtained by varying the spin parameter $j$ from a given fundamental state as in Eq.~(\ref{spinmass}) for which $j$ is an integer in the lab reference frame \cite{Majorana:NC:1932}.
In fact, this family of particles behave at all effects like an infinite spectrum of excited particle states characterized by the angular momentum parameter $j$, as will be discussed in the next chapter too.
When is made such a superposition of Majorana particle states for any value of the angular momentum $j$ each with different values of the phase $\theta_n$ that we can set ad hoc and acceleration parameter $a_n$. Building the superposition we spatially rescaled of a factor $1/a_n$ each component, because all the Rindler observers, instantaneously at rest at time t=0 in the inertial frame, at a given time $t$ will be found at position that depends on the acceleration, $x=1/a_n$,
\begin{equation}
\begin{split}
\sum_{n \geq 1} \frac {e^{i\theta_n} n k_n}{a_n}~K_{\frac 12 \pm \frac{i E}2}\left( \frac m{a_n \left(j+ \frac 12\right)} \right) = \\
\sum_{n \geq 1} \frac {n k_n}{a_n}~K_{\frac 12 \pm \frac{i E}2}\left( \frac m{a_n \left(j+ \frac 12\right)} \right)
\end{split}
\label{superposi}
\end{equation}
here the $a_n$'s build up a set of constant accelerations $\{ a_n\}$ for each of the Rindler particles that we can be suitably chose for our purposes; the parameter $b=\frac{m}{(j+1/2)}$ summarizes the Majorana mass and energy condition and the term $k_n $ is a constant that, multiplied by the integer $n$, corresponds to the chosen acceleration parameter $a$ of the $n$-th state that builds this superposition, $\left( \frac {bn}2 \right)^{-\frac 12 \mp \frac{i E}2} = n k_n/a_n$ for each value of the positive integer $n>0$ and correspondent acceleration $a_n$ of the $n$-th particle.
This can be considered the equivalent of building with a suitable parameter corresponding to our a set of accelerations $\{ a_n\}$ with that of a Dirac-delta potential as discussed in Ref. \cite{sierra}.

The conditions obtained from the energy eigenvalues, imposed to each particle to build an ad-hoc coherent superposition that depends on different values of the Rindler acceleration parameter $a$ of quanta having the same phase factor $\left(e^{i\theta}- 1 \right)$, obtaining an identity as in Eq.~(\ref{majoranarindler}),
\begin{equation}
\left(e^{i\theta}- 1 \right) \sum_{n \geq 1} \frac{n k_n}{a_n}~K_{\frac 12 \pm \frac{i E}2}\left(bn \right) = 0 .
\label{superposi0}
\end{equation}
This is similar to what occurs with a massless Dirac particle. In fact, following \cite{sierra}, the energy-momentum, e.g., of a Dirac massless particle moving in the right, positive x-direction, is given by $p^0 = p^1$ and both are a function of an energy scale, of the relativistic mass and velocity; this suggests that the field theory underlying the Riemann $\zeta$ function, if it exists, must be associated to a massless Dirac or, from our results, to a Majorana particle.

The relevance of Eq.~(\ref{superposi}) in our case is that there exists a correspondence between the modified Bessel functions of the second type $K_\nu(x)$ present in the equation with the Riemann $\zeta(z)$ function based on Kontorovitch-Lebedev integral transforms, as discussed below. This property can be used to set up a correspondence between the zeta function and the superposition of the whole tower states of a given Majorana particle expressed in terms of a series of Macdonald functions $K_\nu(x)$, obtaining, from the identity in Eq.~(\ref{superposi0}),
an integral representation of the coherent superposition of Majorana states discussed before based on the $\zeta(z)$ function.

To better understand this point, of relevant importance is to introduce the topic of the solutions to equations involving both the argument and index of the modified Bessel function of the second kind $K_{\nu} (x)$ (or Macdonald function), which is a well-known issue in the literature, for indices and arguments that take values both in the complex and real domains.

A first example is the solution to the equation $K_{i \nu} (x)=0$, when the index is of imaginary order, $i \nu$, for any fixed argument $x > 0$ is characterized by a set of countably infinite sequence of real zeros \cite{bagirova,Paris}.
In our case, the index $\nu$ is a complex number, $\nu = 1/2 + i E/2$, and,
with this method, this equation has zeros only if the term, e.g., $1/2 + i E/2$ is pure imaginary (as it can occur in decaying or scattering processes involving multi-particle states described by the Majorana Tower \cite{nanni1,nanni2}) and admits an infinite number of zeros, being isomorphic to a one-dimensional Schr\"odinger equation with exponential potential. 
In this particular case, there is in fact a countably infinite number of (simple) real zeros in $1/2 + i E/2$ once $m/a\left(j+ 1/2\right) > 0$ is fixed for the equation 
\begin{equation}
K_\nu(x)=0 
\label{kappanu}
\end{equation}
where the variable is $\nu$ that depends on the angular momentum $j$, for any value of $x$ that depends on the following parameters mass, $m$, acceleration, $a$ and, of course, spin $j$ that characterizes each of the different infinite components that build up the Majorana Tower. 
Interestingly, all these zeros are shown to distribute asymptotically similarly to $\pi(x)$, 
\begin{equation}
\nu_n \sim \frac n{\log n}
\end{equation}
and always depen on the spin parameter $j$ as each imaginary value of $\nu=1/2 + i E_0/(j+ 1/2)$ and $x=m/a\left(j+ 1/2\right)$; in any case, the convergence of Eq.~(\ref{kappanu}) with that of the zeros of $\zeta(z)$ presents similar behaviors.

When the index $\nu$ of the Macdonald function is complex, as in our more general case, one has to find a precise formulation for the search of the zeros of equations like Eq.~(\ref{kappanu}) that, for the specific case when the index is $\nu=1/2 + i \beta$, as in our case (or $\nu=1/2 - i \beta$), and consequently $j$ is real, one can use kernels of modified Kontorovitch-Lebedev integral transforms to solve the equations \cite{rappoport,rappoport1,rappoport2},
\begin{eqnarray}
y_1(x)&=&Re\{K_{\frac 12 + i \beta}(x)\} = Re\{K_{\frac 12 - i \beta}(x)\} =
\\
&=& \int^\infty_0 e^{-x \cosh t} \cosh \frac t2 \cos(\beta t) dt \nonumber
\\
y_2(x) &=& Im\{K_{\frac 12 + i \beta}(x)\} = - Im\{K_{\frac 12 - i \beta}(x)\} \nonumber
\\
&=& \int^\infty_0 e^{-x \cosh t} \sinh \frac t2 \sin(\beta t) dt  \nonumber
\label{rappa}
\end{eqnarray}
that represent two components of the vector function $(y_1,y_2)^T$, solutions of the associated system of Bessel differential equations. 

We do not consider here, for brevity, the more general solutions with $Re\{\nu \} \neq 1/2$ that also include complex values of $j$. We instead to focus our attention to the specific case of $Re\{\nu \} = 1/2$ because we find this term in our equations and also because this particular case is a well known and analyzed topic in the literature of which one can find several analytic formulations.
In this special case, in fact, the Macdonald functions can be calculated with several methods, e.g., by using canonical vector polynomials methods \cite{rappoport3}. More details can be found in Ref. \cite{nist}.

Let us consider the solution with positive energy only for the sake of simplicity. The other term can be easily obtained from the system of equations in Eq.~(\ref{rappa}).
Now, from the Mellin-Barnes integral representation \cite{friot,mellin}, a generalization of Hypergeometric series\footnote{The integral here is calculated, as usual, along a contour obtained from the deformation of the imaginary axis passing to the right of all poles of factors of the form  $\Gamma(\frac 12 + \frac{i E}2)$ and to the left of all poles of factors of the form $\Gamma(\frac 12 - \frac{i E}2)$.} of $K_\nu(x)$ with free parameter $g$ \cite{paris2},
then one writes
\begin{equation}
x^{- \nu} K_\nu(x) = \frac {2^{-\nu} }{4 \pi i}\int^{g+\infty i}_{g-\infty i} \Gamma(s)\Gamma(s-\nu) \left( \frac12 x\right)^{- 2 s} ds
\label{mellinbarnes}
\end{equation}
where $b$ is a free parameter, $g>\mathrm{Max} \{0,\nu\}$, then one finds the identity, by setting, as in our case $\nu=\frac 12 + \frac{i E}2$ as main parameter,
\begin{equation}
\begin{split}
&\sum_{n \geq 1} \left( \frac {bn^p}2 \right)^{-\frac 12 - \frac{i E}2} K_{\frac 12 + \frac{i E}2}(bn^p) = 
\\
&\frac 1{4 \pi i}\int^{g+\infty i}_{g-\infty i} \Gamma(s)\Gamma \left(s- \frac 12 - \frac{i E}2\right) \left( \frac12 b\right)^{- 2 s} \zeta(2 ps)~ds
\end{split}
\label{identity}
\end{equation}
with the parameters $p\geq 1$, $|arg(b)| < \pi/2$, $g > \mathrm{Max} \{\nu, 1/(2p) \}$
and $n \in \mathbb{N}-\{0\}$.
from the hypothesis made for the general solution of Eq.~(\ref{majoranarindler}), if $K_\nu(x)=0$ $\forall x$, $\forall \nu$, which is the condition for the search of a general solution of the equation with variable $\nu$ and free parameter $x$, one can simplify the relation between the modified Bessel function $K_\nu(x)$ setting $p=1$, without loosing in generality, and $b=\frac{m}{(j+1/2)}$, obtaining a link between Eq.~(\ref{superposi}) for a coherent state and the integral of the Riemann zeta function in the complex plane, applying the definition for $k_n$ in Eq.~(\ref{majoranarindler}), 
\begin{equation}
\begin{split}
&\sum_{n \geq 1}~\frac {n k_n}{a_n}~K_{\frac 12 + \frac{i E}2}(b n) = 
\\
&\frac 1{4 \pi i}\int^{g+\infty i}_{g-\infty i} \Gamma(s)\Gamma \left(s- \frac 12 - \frac{i E}2\right)  \left( \frac12 b\right)^{- 2 s} \zeta(2 s)~ds
\end{split}
\label{identity2}
\end{equation}
In this way, writing the identity of Eq.~(\ref{superposi0}) one finds from the Majorana Tower a relationship between the values of the energy $E$ of the particle and thus with the spin $j$ present in the modified Bessel function of the second kind expressed in terms of the Riemann's zeta. 

Rewriting the identity of Eq.~(\ref{majoranarindler}) through Eq.~(\ref{identity2}), with the conditions of a coherent superposition of the Tower states of a Majorana particle, the equivalence obtained from the boundary conditions from the energy eigenvalues for these Majorana particles, with the energies $E$, then becomes
\begin{equation}
\int^{g+\infty i}_{g-\infty i} \left(e^{i \theta}-1\right)  \Gamma(s)\Gamma \left(s- \frac 12 - \frac{i E}2\right)
\left( \frac12 b\right)^{- 2 s} \zeta(2 s)~ds = 0
\label{majrinint}
\end{equation}
with an identity that then actually gives the zeros of $\zeta(z)$, avoiding the identities with zero that may arise if we consider only the trivial zeros from $\left(e^{i \theta}-1\right) = 0$, which can be any function of $s$, different from $s\equiv 0$ that identifies one of the poles of $\zeta$, obtained from the discrete representation of the coefficients in Eq.~(\ref{identity}). The integral representation of a superposition of the tower states of a Majorana particle from Eq.~(\ref{majoranarindler}) and written in Eq.~(\ref{majrinint}) satisfies the energy conditions of the Majorana particle superimposed with its excited tower states when the integral is null. This means that it is null when $\zeta(2s)=0$, namely, when the zeta function goes to zero. This may happen also if the whole integral is null, as a ``global'' behavior of all the paths along the poles, but the result is the residue of the gamma function at the poles usually set for the $\Gamma$'s as $s = - n$ with $n=\{0, 1, 2, ... \}$. To have the whole integral always null then the sum of the residues must be always zero. Another way to obtain this result is through an expansion with Mittag-Leffler functions \cite{mittag} or with other alternative methods that can be found in hypergeometric function analysis that are of use in high energy and particle physics.
 
More precisely, for any value where zeta is null, the energy conditions of the mixture of the tower states of a Majorana particle are satisfied (i.e., when Eq.~(\ref{majrinint}) is null), finding a correspondence with the zeros of zeta and the energy spectrum of a Majorana Tower particle.

Summarizing, we find that the Majorana particles in a Rindler support, as shown in Eq.~(\ref{majoranarindler}), present a relationship between the solutions of their equation of motion and those found for zero mass Dirac particles \cite{sierra}. In a way, the non trivial Riemann zeros (i.e. the imaginary part) are closer to the spectrum of a momentum operator which is first order in the derivatives, like occurs in the Dirac equation, showing a better convergence with zero-mass particles  as occurs for a Majorana particle.

Then, by using the properties of Majorana particles, we find an identity derived from the boundary conditions of the spectrum of energies $E$ of each excited particle state present in the Majorana Tower of a given particle when is cast in a coherent superposition; i.e., a superposition of excited states of the same Majorana particle obtained from the rules used to build the Tower. The integral representation of the energy conditions in Eq.~(\ref{majoranarindler}) results related with the Riemann's zeta through the Mellin-Barnes transformation in Eq.~(\ref{mellinbarnes}, \ref{identity}) and Eq.~(\ref{identity2}), valid for both left and right chirality states, and, more importantly, the energy conditions are satisfied when the Riemann's zeta function goes to zero.

We recall that different chirality states correspond to energies with opposite signatures, as already discussed in the text.
This sets a link between the results of Ref. \cite{sierra}, the Majorana solution and the Riemann Hypothesis: the operator(s) that builds up the boundary conditions of the Majorana particles agrees with the prescriptions of the Hilbert-P\'olya conjecture for both cases with $\nu=1/2+i E/2$ and $\nu=1/2-i E/2$. This because one does not mandatorily require a positive Hamiltonian since the Riemann zeros are positive and negative \cite{sierra0,berry11}.
In a few words, all this ``stuff'' is elementary quantum mechanics and with Majorana particles one does not need to fill in a Dirac sea of negative energy states, unless there is the need of writing and managing additional properties of the physical system like those present in the Majorana solution where energy, mass and momentum are related in a new Hamiltonian as can be found with the other approach we will discuss in the next chapter.


\section{From Hermitian Hamiltonians to the S-matrix method for the RH} 

Another route to the RH is the so-called $S$-matrix method, obtained by managing the amplitudes, zeros and poles of a scattering process, with the standard techniques used to study scattering phenomena also in the presence of infinite-norm vector spaces. 
All the terms used in the previous case, including the already discussed negative energy states, written as $-E$, will appear also here in the analysis of the scattering processes, keeping their meaning, including the properties of the Majorana Tower particles that present a relationship between energy, mass and angular momentum and, being Majorana quanta single particles that coincide with their own antiparticles or superpositions of them.

With the $S$-matrix approach, we will look from a different perspective the results previously obtained with the Majorana quanta by applying the Majorana infinite-components equation in Rindler spacetimes to the quanta described with the tachyonic aspects of the bosonic string.
Then, at the end, we analyze the pole dynamics adopting the analytic continuation in the complex field of the angular momentum $j$ of the Majorana quanta there considered, finding a correspondence between the poles of $S$ and the zeros of the Riemann zeta function in a more general case than that discussed before described by the Macdonald equations, viz., when the index is $\nu_{\pm} = 1/2 \pm i E/2 = 1/2 \pm i E_0/(j + 1/2)$ that implies $j \in \mathbb{R}$ in order to satisfy the conditions of solvability discussed in Ref. \cite{rappoport,rappoport1,rappoport2,rappoport3}.

What one can assume is that if the $S$-matrix approach verifies the RH, there may exist a class of Hermitian Hamiltonians $H$ or other operators $A$ satisfying the Hilbert-P\'olya approach.  
The advantage of using the $S$-matrix approach is that it actually ensures stronger conditions than the other method, asymptotically in the adiabatic regime.  
Of course, this occurs if the idea by Hilbert and P\'olya is valid as we have always assumed, but one cannot exclude that the PT-symmetric Hamiltonians under particular conditions could be the way to prove the Riemann Hypothesis, putting in discussion the validity of all the strict requirements needed for the Hamiltonian $H$ or any other useful operator $A$ to satisfy the requirements for the Hilbert and P\'olya approach.
In other words, if the $S$-matrix method proves that all the zeros of $\zeta(z)$ lay on the critical line, also PT-symmetric systems for which exist at least an operator whose eigenvalue distribution fits with that of the zeros of $\zeta(z)$, can satisfy the Riemann Hypothesis but not vice-versa, if one has to obey to the full requirements of the Hilbert-P\'olya approach.

Obviously, if the $S$-matrix proves that the zeta zeros lay on the critical line, then one can think that there should exist an Hermitian Hamiltonian or a suitable operator describing a particular physical system and satisfying the RH, even if the encouraging results presented in Ref. \cite{bender} with PT-symmetric operators may cast some doubts on the necessity of applying the whole set of strict requirements dictated by the Hilbert-P\'olya approach. A brief discussion can be found in the Appendix (\ref{appA}).

\subsection{The scattering approach to the RH and the Majorana Tower}

Analogies with the scattering amplitude obtained from the scattering matrix $S$ of a physical system and the locations of the zeros of Riemann's zeta have been already discussed in the past, since $1972$, starting with the work by Faddeev \cite{faddeev}, then followed by Lax \cite{lax}, Bhaduri \cite{bhaduri} and many others.

Now, we can put a bridge between the results obtained before with the classical approach by Hilbert and P\'olya and this different approach where one instead searches for the complex poles of the scattering $S$-matrix that can be mapped into the critical line and coincide with the nontrivial Riemann zeroes of $\zeta$ analyizing their properties. In this method, the RH is proved if the $S$ matrix poles and zeros build up a sequence of numbers that overlaps the distribution of the $\zeta(z)$'s zeros and explain the properties of the poles that must distribute along the critical line to be described with the Veneziano amplitude of the tachyonic aspect of the bosonic string scattering already shown in Ref. \cite{castro}.

The main crucial point to solve, to set a path to the proof of the RH here, is related to the fact that the poles of any $S$ matrix used to verify the validity of the RH have the property of existing always in pairs and are always related, in each pair, via complex conjugation.
As stated by Castro in Ref. \cite{castro}, if this property is explained and proved with a physical system, then the proof of the RH with the $S$-matrix approach can be set, drawing the path for a complete route to the demonstration of the RH.

To this aim, we set up our demonstration to this property, by applying the infinite-components Majorana's solution -- the Majorana Tower -- to the study of the zeros and poles of the $S$-matrix of a tachyonic condensate, generated by the scattering of tachyons in the bosonic string theory framework.
Strings and supersymmetric string solutions as superstrings are known to be able to describe states of Majorana quanta, either both bosonic-Majorana, for which $j \in \mathbb{N} - \{0\}$ (pure bosonic, see e.g., \cite{bled,nielsen}, or given by superposition of fermionic states) and the more common fermionic solutions \cite{string}, where, following the original work by Majorana, the angular momentum assumes half-integer values ($j= \frac 12, \frac 32, \frac 52, ... $). As Majorana quanta present different intrinsic (spin) angular momenta for an observer at rest, 
\textit{this family of particles behave at all effects like an infinite spectrum of excited particle states characterized by the angular momentum parameter} $j$. 
For a better insight see the Appendix \ref{appB} and Ref. \cite{tambu2021}.

The quantum system associated to this description is presented in terms of an infinity of virtual resonances described by the corresponding $S$-matrix poles \cite{joffily}, whose differences in the energy levels present similar properties of a system described by random matrix theory (RMT) and reduces the Riemann Hypothesis to an inverse (quantum) scattering problem \cite{khuri}.
Concerning the RMT connection with primes, the RMT statistics apply only over a limited range of energy difference, beyond which there are dramatic deviations. These deviations are precisely those predicted theoretically for a quantum system with a classically chaotic counterpart. 
The RMT universality comes from long classical periodic orbits, whilst the non-universal deviations come from the short orbits \cite{berry175}.

The Majorana Tower in a hyperbolic geometric support, like the Poincar\'e disk or including the case of a general constantly-accelerated reference frame that describes accelerated string ends (see e.g. in AdS/CFT and string and superstring theory \cite{adscft,susskind}), can be successfully used to describe the scattering of two on-shell Majorana tachyons with a Hermitian Hamiltonian \cite{semenov}. 

This can be done because infinite-components field theories like the infinite-components Majorana solutions are known to present ubiquitous space-like solutions and tachyons. These do not only contain massive unitary irreducible representations of the Poincar\'e group for all spins but also the more exotic tachyonic and continuous-spin representations \cite{beka,suda}, with the result of generalizing the spin-energy relationship of Eq.~(\ref{spinmass}) to negative and imaginary values of the spin and energy through the analytic continuation of $j$ in the Regge formulation used to describe the scattering of two (or more) particles \cite{regge,gribovbook}, in a more general approach with respect to that previously discussed with Macdonald functions of index $\nu=1/2 \pm i E/2$. 

We can say that it is the centrality of this relationship in the process of identifying the object-event particle as an inseparable aspect of spin and mass that gives a formidable explanatory force to the Majorana construct in the Regge formalism. All this is already present within the Majorana equation for reasons that we would call ``structural'': the countability-spectrum-mass-spin relationship present in the Majorana solution.
The analytic expansion of the spin variable $j$ has to change according to the evolution of the energy and proper mass in a Rindler spacetime and then becomes a function of the square root of the energy of the exotic state, with a linear dependence from the energy of the particle in its fundamental state, Poincar\'e transformations including Lorentz boosts.

In the Majorana solution one can build a set of exotic states through an infinite sum of four-spinor operators. 
In this way, the system can describe any interaction between different types of particle solutions resulting from the combination like a condensate of particles traveling slower, equal and faster than light (bradyons, tachyons and luxons), as described in \cite{nanni1,nanni2}, in a field consistent with the CPT invariance. 
This is in total agreement with the main properties of Majorana infinite-components quantum fields that include spin, mass distribution and tachyonic solutions \cite{gu,aste,casalbuoni} where the basic property of a Majorana particle is that a particle state, with its definite four-momentum and spin, can be transformed by a CPT transformation and a subsequent (space-time) Poincar\'e transformation into itself.

\subsection{$S$-matrix tachyonic bosonic string and the Majorana Tower}
After this introductory part to the $S$-matrix, we now use the Majorana Tower to give the needed physical interpretation and justification to explain the some still not clear properties of the amplitudes of the scattering matrix $S$ describing tachyon scattering phenomena in the framework of the bosonic string theory that are deeply related with the RH in a hyperbolic geometry (like the Poincar\'e disk) like the \textit{s}-wave scattering. Here, the nontrivial zeros and the complex poles of the $S$-matrix correspond when the condition $z_n = 1/2 + i \lambda_n$ is satisfied.
In this way, a Wick rotation of the critical line of the nontrivial zeros of $\zeta(z)$, would overlap with the poles of the scattering matrix, given by the double zeros at the denominator.

Following the claim in Ref. \cite{castro}, with the tachyonic solution of the Majorana Tower already discussed by Majorana in 1932, we find an explanation for why in this physical systems for which the poles of the $S$ matrix lay on the critical line, the complex double poles of the $S$-matrix always occur in complex-conjugate pairs,
\begin{equation}
z_n = 1/2 \pm i \lambda_n.
\label{zeros}
\end{equation}
From $S$-matrix theory, as usually occurs in a Hermitian system, the poles and the zeros of the scattering matrix $S$ always are presented and exist in pairs and their relationship occur via complex conjugation.
This would imply that the poles of the $S$-matrix overlap the Wick rotation
of the nontrivial Riemann zeta zeros. This represents the required physical proof to validate the Riemann Hypothesis.

The link between the Majorana solution and string theory for the RH starts from the formulation that derives from the QM of a particle moving in the hyperbolic plane \cite{fadeev} and from a clear relationship existing between the bosonic and more general formulations of String Theory and Higher Spin fields, exactly like those already present in the Tower solution \cite{sagnotti}.

To this aim, one has to fix better this point by analyzing a tachyon scattering process. When a tachyon is produced in the s-channel by the scattering of two incoming tachyons, it decays into the tachyonic condensate state made out of two tachyonic-resonances, as described in bosonic string theory. 
The multi-particle tachyonic scattering finds deep links with the RH, as the complex-conjugated double poles of the S-matrix found satisfy the CPT condition for tachyonic scattering \cite{castro,he15}.

The required conditions for the RH can be fulfilled by a tachyonic set of particles described by the family of Majorana quanta.
In the bosonic string approach, the location of the nontrivial zeros of $\zeta(z)$, as in Eq.~(\ref{zeros}), were proved to correspond to tachyonic resonances, namely, tachyonic condensates that are associated with the bosonic string scattering amplitudes through the analysis of the s-channel of the scattering matrix $S$.
The real part of the Riemann zeta zeros has been shown to admit only the value $1/2$ for their real part. 

The use of the $S$ matrix implies the adoption of a certain number of constraints: block decomposition, Lorentz invariance, analyticity (causality), crossing and unitarity. The Mandelstam analysis shows that if the exchanged particle is stable the pole is on the real axis, otherwise it is at a distance from it inversely proportional to its average life.
In the relativistic case the Regge trajectories are linear (almost perfectly), and the Virasoro (Euler) formula reveals a complete crossing symmetry in these structures. 
The s-wave amplitudes (s-channel) of the scattering matrix $S(k)$ that describes the scattering of a particle in the Poincar\'e disk or in a hyperbolic plane 
\begin{equation}
S(k) = \frac{\zeta(i k)\zeta(1 - i k)}{\zeta(1 + i k)\zeta(- i k)}
\label{smatrix}
\end{equation}
has complex poles that correspond to the zeros of $\zeta(z)$ in the denominator and to their poles in the numerator. 

The key to the RH is on the scattering resonances  given by the complex poles of the S-matrix that have been shown to be located on a horizontal line below the real axis and can be mapped into the critical line of zeros of the Riemann zeta function. 

In the case of the hyperbolic plane, particles with s-wave scattering have the complex-poles of the S-matrix that correspond to the nontrivial $\zeta(z)$ zeros if they satisfy $k_n = i (1/2 + i \lambda_n)$ that, followed by a Wick rotation of the Riemann critical line, would yield the complex momenta of the scattered tachyons be related with the double poles of the S-matrix \cite{castro,schumi}.
These complex momenta are the complex double poles of the scattering matrix are given in complex-conjugated pairs $ - i k_n = (1 + i k_n)^*$ and $k_n = i (1/2 + i \lambda_n)$.

Of course, the values of $\lambda_n$ may vary depending on the set of observers, e.g., if one describes the motion of the particles from the point of view of a family of Rindler-like hyperbolic spacetimes or if one describes the scattering process of Rindler uniformly accelerated particles in the lab reference frame. At all effects, the sequence of $\lambda$'s is given by the poles of $S$ that depend from a coordinate transformation.
As usual, the poles of $S$ can be located in the lower half-plane, describing mode exponential attenuation, while the zeros are found in the upper half-plane and give the mode divergence.

What can be shown is that the imaginary values $i \lambda_n$ that are a function of the energy, in the Regge formulation and with the Majorana relationship between energy and angular momentum can overlap those of $\zeta(z)$ thanks to the dynamics of the particles and of the geometric properties of the hyperbolic support of the Rindler spacetime. 

Noteworthy, if one considers the acceleration parameter $a$ as a free parameter or a function of time, the spin parameter $j$ varies continuously (as already discussed by Majorana in his original work) and one can in principle fit the distribution of the poles with that of the zeros of $\zeta(z)$, as shown by adopting the prescriptions of the Regge trajectories obtained from the conservation of the angular momentum in hyperbolic spaces \cite{lavenda}. In fact, Regge trajectories relate bound states like the Majorana Tower particle with $j=0$  with the resonances having the same quantum numbers that in the Majorana Tower are the same particle with mass $m$ but with different values of the angular momentum $j$, i.e., different values of the Majorana mass, $M^*=\frac m{j+ \frac12}$, with which we obtain, for this type of particle, an algebraic equation of the third degree in $j$ that describes the effect of an hyperbolic spacetime on the Regge trajectories,
\begin{equation}
j^3 + \left(1- \frac{m}\omega\right) j^2 + \left(\frac14 - \frac{m}\omega\right) - \frac{m}{4 \omega} - \frac{\Lambda m}{\omega} = 0 \, ,
\label{reggetower}
\end{equation}
which gives two complex-valued solutions of the angular momentum $j$, as expected from the Regge theory, and one real solution if the parameters satisfy the following condition, 
\begin{equation}
\begin{split}
&4 \left(\frac14 - \frac{m}\omega\right)^3-\left(1- \frac{m}\omega\right)^2 \left(\frac14 - \frac{m}\omega\right)^2  - 
\\
&- 4 \left(1- \frac{m}\omega\right)^3 (\frac{m}{4 \omega} + \frac{\Lambda m}{\omega}) + 18 \left(1- \frac{m}\omega\right) \left(\frac14 - \frac{m}\omega\right) \times
\\
& \times (\frac{m}{4 \omega} + \frac{\Lambda m}{\omega}) + 27 (- \frac{m}{4 \omega} - \frac{\Lambda m}{\omega})^2\geq 0,
\end{split}
\label{jreal}
\end{equation}
that can be used in also in the previous approach with the Macdonald equations $K_\nu(x)$.
The parameter $\Lambda$ is the scale parameter, $\omega \sim \sqrt{a}$ represents the angular velocity of the rotation of a uniformly rotating rigid disk or the rotation of the string that represents the Majorana states that imposes an hyperbolic structure of the spacetime correspondent to a Rindler spacetime with acceleration $a$. In fact, Regge trajectories can be used to describe relativistic rotational states of the particle trajectory that are quantized by their values of $\Delta j$, favoring with these rules the search for the zeros of Eq.~(\ref{kappanu}) and thus the zeros of Riemann's zeta function from the identity of Eq.~(\ref{identity2}), as required by the boundary conditions for Majorana Tower particles expressed in Eq.~(\ref{majoranarindler}).

The reason for the successes of the formalism here used for the RH based on Regge and Veneziano works is that the latter ones are linked to the fact that they adopted a non-perturbative approach, which highlighted a structure that theoretically it could have been found many years ago. In other words, their analysis is valid and functional even before merging into string theory. What was missing was a construction like the Tower,  as it is purely relativistic, which acts as a frame that selects the spectrum and avoids ghost states (a Goddard-Thorn no-ghost theorem \cite{goddard} is not necessary). That is, it acts as an ante litteram ``holding up'', giving the values with the amplitude that goes as $1/2$ but also the correlated values due to the centrality of the mass-angular momentum relationship. This allows Regge's relations to be applied also to different situations, such as the Recami's spectrum of black holes \cite{recami}.

To discuss the connections between the Riemann's zeta zeros and the amplitude and poles of the $S$-matrix, as in Ref \cite{castro}, we introduce the Mandelstam variables for each different channel, $s = (k1 + k2)^2$, $t= (k2-k3)^2$, $u= (k1-k3)^2$ and the Regge trajectories in the respective $s$, $t$, $u$ channels
\begin{equation}
-\alpha (s) = 1 + \frac 12 s, \,\, - \beta (t) = 1 + \frac 12 t, \,\, - \gamma(u) = 1 + \frac 12 u,
\end{equation}
then we write the Veneziano amplitude \cite{veneziano} through the Regge trajectories $\alpha$, $\beta$ and $\gamma$
\begin{equation}
A(s,t,u) = \frac{\zeta(1-\alpha)}{\zeta(\alpha)}\frac{\zeta(1-\beta)}{\zeta(\beta)}\frac{\zeta(1-\gamma)}{\zeta(\gamma)}
\label{venessia}
\end{equation}
where $\alpha = 1/2 + i \lambda$; then one can determine the values of $\beta$ and $\gamma$ choosing the momenta values for the $s$, $t$, $u$ Mandelstam variables, $k_1$, $k_2$ and $k_3$, $k_4$, to obey the tachyonic on-shell condition associated with the ground state of bosonic open strings given in a hyperbolic support like the Poincar\'e disk or the Rindler spacetime.

In this case, the four-point dual bosonic open string, defined in a hyperbolic support, is known to provide a clear physical interpretation in terms of tachyonic scattering described by the poles $k_n$, overlapping the distribution of the non-trivial $\zeta(z)$ zeros. 

The Majorana Tower can actually provide the missing link to complete the picture for a road to the RH, namely, the searched physical interpretation necessary to explain the outcome in complex conjugate pairs of the amplitudes and corresponding poles of the $S$ matrix. This can be shown with a simple example involving Majorana tachyons.
In the original work by Majorana, the first type of tachyonic state is obtained from the following energy-momentum relationship that, for any given value of the momentum $p$ of the particle gives a couple of imaginary values of the energy with opposite signs,
\begin{equation}
E^2 = p^2 - m^2, \,\,\, p^2 < m^2 ,
\label{tachyonic}
\end{equation}
with real mass $m$ but imaginary energy values $\pm iE$. It is well known that the imaginary parts of the energies in scattering theory correspond to the inverse lifetime of particle-resonances. The resonance-width is the inverse of the lifetime.

Thus, the mirror solution to the pole associated to the positive imaginary value $iE$ is that corresponding to the complex-conjugated pole characterized by the negative imaginary energy value $- i E$. These two values correspond to the two independent particle solutions to Eq.~(\ref{majarray}) that actually behave as two independent Majorana particles characterized by different chiralities $L$ or $R$, because as already said, in the Majorana solution any antiparticle state has to coincide with its particle states. An example is that previously discussed in Chap. II, in a $D=1+1$ Rindler spacetime where different chirality states correspond to the propagation of the two Majorana particles along the left or right directions with respect to the origin, as in this special case mirror reflection coincides with chirality as shown and discussed in Ref. \cite{zra}.
The other exotic types of tachyonic particles with imaginary mass and also negative rest mass, possible solutions present in the Majorana solution represent a class of solutions that is not now of interest to us.

Consider now the scattering of two Majorana tachyons having the same rest mass $m_1=m_2 = m$ and momenta $k_1=k_2=k$. We apply the rules for the scattering used in the bosonic string to find that the imaginary part of the poles $k_n$, corresponding to the real part of the zeros of $\zeta(z)$, has to be $1/2$, as expected by the RH. To this aim, one has to assume the following, already known, condition for the Mandelstam variables
\begin{equation}
s+t+u=-8 
\label{onshell}
\end{equation}
with the on-shell condition $s=-2$, $j=0$ and $k_i = -2$, for the ground state as in the case of the open bosonic string. 
If this condition is not satisfied what we search cannot be poles of the Veneziano amplitude $A(s,t,u)$, to obtain the whole spectrum of $\lambda_n$ now one has to keep valid the constraint in Eq.~(\ref{onshell}) and vary the momenta of the particles. Thus the poles are located across the critical line. This means that the searched value must be $1/2$ as proved in Ref. \cite{castro}.

If the particle is a Majorana-Tower particle, we can have additional properties with respect to the bosonic string that could be useful for our purposes, namely a relationship between energy and mass involving the angular momentum $j$.
When one calculates the Regge trajectory $\alpha= 1/2 + i \lambda$, imposing the on-shell condition of the Mandelstam variable $s = (k1 + k2)^2 = k_1 k_2$, then obtains $- \alpha(s)= 1+ 1/2s = 1+ 1/2 k_1 k_2$.
Written in terms of the Majorana energy and mass and assuming $c=1$, then one obtains the following relationships for the s-channel  
\begin{eqnarray}
&s& = \pm \left(\frac m{j+\frac12} \right)^2 \left(j^2+j-\frac 32 \right) \nonumber
\\
- \alpha&(s)& = 1 \pm \frac 12 \left(\frac m{j+\frac12} \right)^2 \left(j^2+j-\frac 32 \right)
\label{alfas}
\end{eqnarray}
where $j$ is the Majorana spin with the tachyonic condition of Eq.~(\ref{tachyonic})
already discussed in the 1932 work by Majorana that leads to $p^2=E^2 + m^2$ with real-valued mass tachyonic particles.
The s-channel has also to satisfy the constraint dictated by the hyperbolic geometry of the spacetime expressed by the Regge equation in Eq.~(\ref{reggetower}), where the parameter $\omega$ corresponds to the local Rindler acceleration parameter $a$ via $a \sim \omega^2 r$, where $r$ is the distance from the center of rotation (in a simplified newtonian framework). Anyway the relationship $\omega \propto \sqrt{a}$ keeps its validity also in the weak relativistic regime. For a better insight towards the relativistic approach see \cite{jonsson,cohen}.

\begin{figure}[!htbp]
	\begin{center}
	\includegraphics[width=1.05\columnwidth, height=0.7\columnwidth]{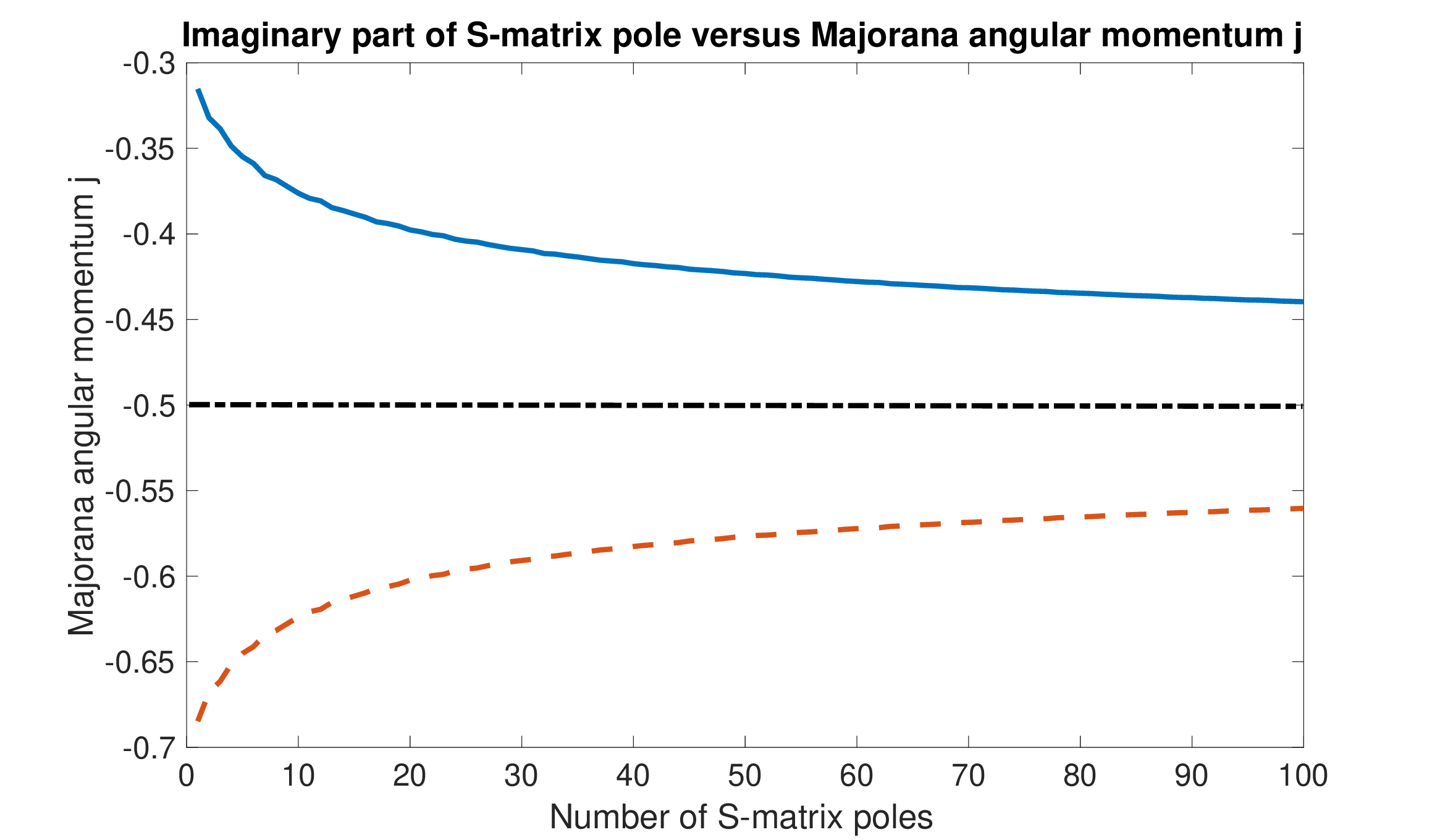}
	\caption[]{Distribution of the first $100$ angular momentum values $j$ with respect to the imaginary values $- i \lambda_n$ of the poles of the $S$-matrix that show how the zeros behave. The distribution occurs in pairs as expected from the scattering matrix zeros/poles, reinforcing this route to the RH as in Ref. \cite{castro}. The behavior is quadratic convolved with a log-type distribution, describing a uniformly accelerated particles in an anisotropic universe without matter with quadratic behavior. The symmetry axis is the horizontal line $y=- 1/2$. The value of the rest mass of the Majorana quanta, $m$, is an arbitrary constant.}
	\label{fig1}
	\end{center}
\end{figure}

The properties of the hyperbolic spacetime, imply three constraints for $j$ in the Regge trajectories as a function of the Rindler acceleration parameter $a$, two are complex-valued and one real if the condition in Eq.~(\ref{jreal}) is satisfied.
After that, from Eq.~(\ref{alfas}) one recovers the values of the imaginary part of the pole by varying $j$ in its analytic continuation, 
\begin{equation}
- i\lambda_j = \frac 14 \left[ \frac12 \pm \left(\frac m{j+\frac12} \right)^2 \left(2j^2+2j-3 \right)\right]
\label{ilambda}
\end{equation}
from the Majorana spin-mass relationship of Eq.~(\ref{spinmass}) and the Regge equation of Eq.~(\ref{reggetower}), one gives a physical explanation in terms of $j$ to the reason why the amplitudes of the scattering matrix $S$ always give poles in complex-conjugated pairs, but for a constant, as shown in Fig. \ref{fig1}: from the values of the poles one obtains the values of the angular momentum $j$ that is distributed in pairs, centered on the $y=-1/2$ axis due to the definition in Eq.~(\ref{spinmass}) and from the interpretation of Majorana particle and antiparticle states obtained from Eq.~(\ref{majarray}) already discussed.

There is an important issue one has to pay the attention to and understand that the analytic continuation of $j$ is already present in the original work by Majorana without introducing ad-hoc properties to explain what has been discussed: only in the lab reference frame the Majorana Tower is made with particles carrying discrete in spin, characterized by integer or half integer spin values. As written also in his 1932 original work, by changing the reference frame energy changes and thus the angular momentum $j$ is not mandatorily integer, even if the spectrum is discrete an infinite, as shown experimentally in Majorana quasiparticle states used for quantum computation, the so-called Majorana anyons \cite{anyon}. 
The behavior observed in the angular momentum recalls that of a uniformly accelerated particle in an anisotropic universe without matter, described by a Rindler spacetime with a quadratic-type Kasner metric \cite{kasner,rindler,mtw}. As said before, the real part of the pole must be always $-1/2$, in agreement with the RH, otherwise the scattering is not governed by the Veneziano Amplitude in Eq.~(\ref{venessia}), going against the initial hypotheses.

The discrete number of the imaginary parts of the nontrivial $\zeta(z)$ zeros and of the $S$-matrix poles, $\lambda_n$, can be also interpreted geometrically in terms of a discrete set of numbers that gives the distances $d_n$ between the two variable vertices of an Euclidean triangle identified by three vertices that degenerates into a vertical strip in the upper complex plane and the other vertex is located at infinity. 

Let us consider, as discussed in \cite{castro}, the case where $\alpha +\beta = 1$ and $\alpha + \beta +\gamma = 1$, i.e., for $\gamma = 0$, then one obtains a  discrete sequence of real numbers,
\begin{equation}
d_n= \frac \pi{\cosh(\pi \lambda_n)}
\label{dienne}
\end{equation}
that describes the distribution of the scattering poles in relations that involve analyticity, the Veneziano bosonic string amplitudes, the associated Regge trajectories, and M\"obius transformations. The Schwarz-Christoffel and Riemann-Schwarz functions map the upper half s-plane onto a triangle. This makes possible the connection of the Regge trajectory functions to the generators of the underlying invariance group of discontinuous transformations \cite{Choudhary}.
When one inserts the values of Eq.~(\ref{ilambda}) obtains a complicated function of the angular momentum $j$,
\begin{equation}
d(j)= \frac \pi {\cosh\left\{-\frac \pi4 \left[ \frac 12  \pm \left(\frac m{j+\frac12} \right)^2 \left(2j^2+2j- 3 \right)\right]\right\}}
\label{diennecont}
\end{equation}
with which one gets the values of $j$, following Regge theory, that at the end correspond to the poles of the scattering matrix $S$. 

From this, the RH seems to be satisfied when one assumes the hypotheses made in Ref. \cite{castro} and applied to the family of Majorana particles in the Tower as they behave as excited states of the main particle with the lowest angular momentum ($j=0$), providing, with the dependence of the energy $E$ from $j$, a physical interpretation that is needed to close the loop in this novel approach to the Riemann Hypothesis.

The possible proof of the Riemann Hypothesis is based on the existence of a special type of poles of the open bosonic string scattering having Veneziano amplitude $A(s, t, u)$ and corresponding to the nontrivial zeta zeros, with $Re(Z) = 1/2$, and have a unique correspondence to a tachyonic pole in the $u$-channel with mass-squared $-2 m^2_P=-2$ and zero angular momentum, $J = 0$, if and only if, $\alpha = \beta^* = 1/2 + i \lambda$ and $\gamma = 0$ and one obtains the remaining things by using cyclic symmetry on the other channels.

The location of the Riemann critical line of zeta zeros given by the complex numbers $\alpha = 1/2 + i \lambda$ does correspond to a real valued pole of the scattering amplitude $A(s, t, u)$ when are assumed complex-valued energy-momenta and angular momenta. In the Majorana Tower framework, the omnipresent mirror solution has a clear physical explanation in terms of two independent Majorana quanta with energies $\pm E$, and for each chirality state each particle coincides with its antiparticle, this giving the physical explanation of the existence of the two complex-conjugated poles we were looking for.

\subsection{$S$-matrix pole dynamics}
\textit{``Just one more thing''}, as lieutenant Columbo was used to say before cornering someone with an inescapable line of questioning as made here, \textit{``there are a couple of loose ends I'd like to tie up''} \footnote{The Lieutenant Columbo is the main character of the famous  American crime drama TV series starring Peter Falk, a homicide detective from the Los Angeles Police Department.}: 
it is interesting now to give a physical description or, as we can say, an interpretation of what has been discussed up to now developing a dynamics of the poles $k_n$ of the $S$-matrix based on the hyperbolic spectrum of the $d_n$'s with the analytic continuation $d(j)$.
These numbers that actually represent the regularized values of the Veneziano amplitude $A(s, t, u)$ can describe the physical system from another perspective: what is shown in Eq.~(\ref{dienne}) it can be used to describe our dynamical system in terms of an other system based on the dynamics of the $S$-matrix poles. In our case the $S$-matrix can be solved exactly and the pole dynamics reflects the dependence on $j$ of the dynamics of our system.

Generally, the pole dynamics occurs in the complex plane; the pole locations, and their flow and motion in the complex momentum plane that can be generated by the deformation of the potential or by variations of the scattering parameters are of great interest to characterize the physical system \cite{belchew,nuss}.
In the complex plane, as long as two conjugate poles exist, their trajectories have to be symmetric with respect to the imaginary axis.

With the way we selected the $d_n$ variables and thanks to the hypotheses we made with the Veneziano amplitude $A(s,t,u)$, this implies that the imaginary part of the pole $k_n$ is always $1/2$, corresponding, after a Wick rotation, to the real part of the correspondent zero $z_n$ of the Riemann's $\zeta(z)$ in agreement with the RH. Thus, the dynamics will be confined in the domain of real numbers when the pole coincides with a zero of zeta and when has to be in the critical line. They are related to the zeros of the Jost function for cutoff potentials in the complex momenta plane in a one-to-one correspondence as the $S$-matrix is defined through the Harish-Chandra $c$-functions, and the Jost functions describes the scattering in the Poincar\'e disk
\begin{equation}
S=\frac{c(k)}{c(-k)}
\end{equation}
through the $S$-matrix, with the same meaning as in Eq.~(\ref{smatrix}).
This also excludes other type of poles that are not associated e.g. with normalizable states, and so on \cite{newton}, and the energy and width ratios of the resonances result associated to the nontrivial zeros of $\zeta(z)$  \cite{Joffily2}, implying that there are not other types of poles in the dynamics we are considering.

The connection with the original dynamical system is interesting and quite straightforward as the $d_n$'s are given by the Regge trajectories $\alpha$, $\beta$ and $\gamma$ that depend on the momenta $k_i$ of the particles in the scattering process, namely, from the actual properties of our system and on the angular momentum $j$ because of the Majorana Tower.

To explain in a few words the link between our system and the pole dynamics, let us start from $S$-matrix formulation of the bosonic string or equivalently of multi-spin states described by the Majorana Tower; one can associate these properties again to those of the Majorana Tower in a Rindler $D=1+1$ spacetime. In this hyperbolic system, because of the scattering properties already discussed, the amplitudes and poles are given always in pairs and related via complex conjugation.
This is nothing but a mathematical transformation obtained using a biunivocal linear transformation which links any previous (old) state vector with the new one through e.g. a square nonsingular matrix. In this way, one obtains a different but equivalent mathematical description of the physical system where, the imaginary part, $\lambda$, related to the acceleration $a$ in the local reference frame also depends on the angular momentum $j$ of the corresponding Majorana solution as described by the Majorana Tower in its analytic continuation. We actually describe the Majorana quanta in the condensate produced by the scattering whose poles are given by Eq.~(\ref{venessia}) with the on-shell conditions of Eq.~(\ref{onshell}) and identified by Eq.  (\ref{dienne}).

The dynamics of the poles is connected with the infinite and discrete number of the imaginary parts of the nontrivial $\zeta(z)$ zeros. The $S$-matrix poles, $\lambda_n$, can be interpreted geometrically in terms of a discrete set of numbers that gives the distances $d_n$ between the two variable vertices of an Euclidean triangle identified by three vertices that degenerate into a vertical strip in the upper complex plane and the other vertex is located at infinity. 
In this way, one can set up the $S$-matrix pole dynamics in the hyperbolic spacetime $(\eta,\xi)$ and find further connections through the distribution of the $S$-matrix amplitudes and poles of the dual bosonic string scattering that describes the tachyonic condensate considered in the scattering process in a Rindler spacetime with the Majorana Tower.

This approach recalls the main ideas that initiated $S$-matrix program, as it is a non-perturbative analytic approach to the scattering problem in quantum field theory. The main idea was to determine only the scattering amplitudes and mass spectra of a given physical system without using the Lagrangian formulation, obtained by only imposing analytic constraints on the scattering matrix $S$. In this way, the $S$-matrix becomes the operator that relates initial and final states in a scattering process, and then one has to provide a clear physical interpretation to all the singularities that characterize the process.
Anyway, as our system is $1$-dimensional, being a hyperbolic $D=1+1$ spacetime, from the Coleman-Mandula theorem, the knowledge of the amplitudes for the scattering between two-quanta states is sufficient to describe the system \cite{cm,bomba}.

The link is obtained by taking the geometrical description of the $S$-matrix zeros in term of Euclidean triangles 
and use the distance terms of each associated Euclidean triangle related to the $\lambda_n$ terms to describe the dynamical properties of the Majorana particles of the tachyonic condensate in the $D = 1+1$ Rindler spacetime.
Consider the Rindler coordinate system $(\eta, \xi)$ obtained from the Minkowski coordinates $(t,x)$,
\begin{eqnarray}
t&=& \frac{e^{a \xi}}a \sinh(a \eta)=\frac{e^{a \xi}}a \sinh(\pi \lambda_n)
\\
x&=& \frac{e^{a \xi}}a \cosh(a \eta)= \frac{e^{a \xi}}a \cosh(\pi \lambda_n) \nonumber
\label{rindler}
\end{eqnarray}
if we set $\lambda_n = a \eta/\pi$ we define a new hyperbolic coordinate system related to the poles of the $S$-matrix and its related amplitudes,
\begin{eqnarray}
t&=& \pm \frac{e^{a \xi}}a \sqrt{\frac{\pi^2}{d_n^2}-1} \nonumber
\\
x&=& \frac{e^{a \xi}}{a} \frac {\pi}{d_n}
\label{rindler2}
\end{eqnarray}
with the forward $(+)$ or backward $(-)$ in time solutions. 
When one considers the Majorana particles of the Tower, varying the value of the angular momentum $j$, they are interpreted as an excited state of the particle with the lower angular momentum in agreement both with Majorana and Regge theories.
Then one can build a point to point ($1-1$) mapping function $f$ \cite{lacasa} that maps the zeta's zeros to the representation of the scattering poles of $A(s,t,u)$ given in terms of the distance of the two vertices in Euclidean triangles with the parameter $d_n$, to the dynamics in the hyperbolic spacetime defined by this relation, taking in account the Rindler metric $(\eta,\xi)$ and the acceleration $a$ 
\begin{equation}
f: z_n=\frac 12 + \lambda_n  \xrightarrow[\text{$j$ Majorana}]{\text{$S$-matrix}} d_n \xrightarrow[\text{Rindler}]{\text{Majorana}} d(j)_{n \equiv j} .
\end{equation}
In this way one obtains a set of angular momenta and coordinates corresponding to the zeros of zeta as a function of the angular momentum $j$ and parameter $a$ in the hyperbolic spacetime $(\eta,\xi)$. The angular momentum $j$ then introduces a quadratic term that can be explained as a convolution with a Kasner spacetime. The parameter $a$ can be used to better fit the poles and the asymptotic behavior of the dynamics can be tested by using the number variance of the zeta zeros in Ref. \cite{berry175}.

Following the initial ideas by Majorana and Regge, when one extends these equations to the continuum for the pole dynamics with $d_n\in \mathbb{R}$ or $d_n\in \mathbb{C}$, one then calculates the momentum of this dynamics,
\begin{equation}
p = \frac{m}{\left( j+\frac 12 \right)} \frac{e^{2 a \xi} \dot{\xi}}{\sqrt{e^{2 a \xi}(1-  \dot{\xi}^2)}}
\label{hmomento}
\end{equation}
and the Hamiltonian
\begin{eqnarray}
H &=& \frac{m}{\left( j+\frac 12 \right)} e^{a \xi} \sqrt{\frac{e^{2 a \xi} \dot{\xi}^2}{(1-  \dot{\xi}^2)} \frac{\left( j+\frac 12 \right)^2}{m^2} + 1} 
\\
&=& M^{*} e^{a \xi} \sqrt{\frac{e^{2 a \xi} \dot{\xi}^2}{(1-  \dot{\xi}^2)} \frac 1{M^{*2}} + 1} \nonumber
\label{acca}
\end{eqnarray}
that, with simple algebra, it can be shown that also this belongs to the class of $xp$ Hamiltonians. The term $M^{*2}$ is the Majorana mass shown in Eq.~(\ref{majmass}).
When $j \in \mathbb{R}$, the Hamiltonian is Hermitian and describes the pole -- or a migration of them -- in the real plane representing the behavior of a Majorana tachyon obtained when is satisfied the tachyonic condition in Eq.~(\ref{tachyonic}) and by assuming $M^{*} \in \mathbb{R}$. 

This means that, as in the case studied in Chap II, $j$ must be real and obey the condition dictated by the Regge trajectories in a Rindler or more general hyperbolic spacetime written in Eq.~(\ref{jreal}). This occurs when there is correspondence between the poles $k_n$ and the non-trivial zeros $z_n$ of $\zeta(z)$ that satisfies the Hilbert-P\'olya formulation of the RH having the trivial zeros with real part $1/2$.

In any case, if there were nontrivial zeta zeros outside the critical line, violating the Riemann hypothesis, i.e., hypothetical zeros with real part $\mathrm{Re} (z) \neq 1/2$, these hypothetical zeros -- even if inside the critical region and discussed with the Majorana solution -- would not correspond to any of the zeros of $A(s,t,u)$ violating the initial hypotheses assumed in our system and the $d_n$ or $d(j)$ are not real any more.

To conclude, we present some additional interesting analogies with different physical systems that are mathematically described by the Majorana Tower.
The Majorana Tower finds other interesting relationships with the RH that are not limited to scattering phenomena only: when the internal motion of a system can be described by algebraic methods, the equation of motion is equivalent to a Majorana equation and the Regge poles represent the stable solutions. One example is the Bethe-Salpeter equation of a composite system with non-relativistic internal motion describing the singularities of the scattering amplitude of an Hydrogen atom that presents an $SO(4,1)$ spectrum-generating algebra and, for an observer in motion, the system is described by an $SO(4,2)$ algebra. This algebra contains the generators of the Lorentz group with a correspondence to the infinite-components Majorana solution \cite{bisiacchi67}. The energy levels appear to be distributed similarly to the non-trivial zeros of $\zeta(z)$ \cite{iliev} as occurs with random matrix theory \cite{bee}.
Noteworthy, the description of a particle in the Rindler spacetime is isomorphic to the spectrum of an hydrogen atom \cite{khuri}, and thus to the Majorana Tower, together with the connections with the zeros of $\zeta(z)$ and for the fermionic solution. 
Real examples of Majorana Tower solutions have been actually experimentally observed with quasiparticles made with vortices of light \cite{tambu2021} and in resonant plasmas \cite{Tamburini&al:EPL:2010,TowerTT}.

\section{Discussions and Conclusions}
Hilbert, Hardy and P\'olya made this treatment of the RH moving towards the study of dynamical systems. From the results obtained with the Hilbert-P\'olya approach we passed to the $S$-matrix formulation to treat the physical analogue of the RH in a complete and coherent way. The reason is actually explained by the Majorana Tower which is the perfect expression of a ``Reggeized'' world that does not contain a Higgs-like mechanism and does not know other perturbative processes.

We applied the infinite-components Majorana solution to these two approaches to verify the validity of the Riemann Hypothesis by using the results in Ref. \cite{sierra,castro} where we think that a pathway to the RH was clearly shown.

The first one is that proposed by Hilbert and P\'olya, where the RH is true if there is a an Hermitian or unitary operator, describing a given physical system, whose eigenvalue spectrum has a $1-1$ correspondence with the non-trivial zeros of $\zeta(z)$, even if the results in Ref. \cite{bender} obtained with PT-symmetric Hamiltonians in Rigged Hilbert Spaces can in principle cast some doubt on the strict requirements of Hermiticity or self-adjointness present in the original formulation of the Hilbert-P\'olya approach.

Here, after having analyzed the various attempts made in the literature, we proposed the Majorana family of particles given by the Majorana Tower as further development to the studies made with the Dirac particle. We found that the Majorana particles present in their equations of motion in a Rindler spacetime a distribution of energy values, as a function of the angular momentum $j$ that agrees with that of the zeros of $\zeta$, realized with an analytic approach through Kontorovitch-Lebedev integral transforms and second order modified Bessel functions of the second kind derived from the boundary conditions for Majorana particles in a $D=1+1$ Rindler hyperbolic spacetime, described by an Hermitian operator. We find an effective correspondence with the Riemann $\zeta(z)$ function through an implicit equation involving an integral transformation and with the zeros of zeta.

The second method is instead based on the correspondence between the non-trivial zeros of $\zeta(z)$ and the zeros and poles of the $S$-matrix of scattering Majorana tachyons and this novel approach to the RH.
The trivial zeros of $\zeta(z)$ correspond to the infinite countably poles of the Veneziano amplitude found in the negative real axis through the immediate correspondence with the trivial zeta zeros that lie in the negative even real axis, viz.,
$- n \rightarrow - 2 n$.

From the literature one can show that it is sufficient to prove and justify with a physical model that the poles of the scattering matrix $S$ are always given with the form $k_n = 1/2 \pm i \lambda_n$. This is our contribution obtained by applying the properties of Majorana tachionic solutions to the bosonic string.
The key issue in this approach, to prove the RH, is to find not only a distribution of poles that fit that of the zeros of $\zeta$ but it needs a clear explanation why the poles are given in pairs and result related through complex conjugation. 
In this case we not only characterize the zeros of $\zeta(z)$ through the poles and zeros of the scattering matrix with the Veneziano amplitude $A(s,t,u)$ as already discussed in the literature, but provide a physical justification to this procedure through the properties of the Majorana solution, indicating that the real part of the zeta zeros, corresponding to a Wick rotation of the poles of the scattering matrix is at all effects $1/2$, in agreement with the RH.

As we have shown, the Majorana infinite-components solution can provide the required physical explanation for which the amplitudes of the scattering matrix of the bosonic string in terms of tachyonic Majorana particle-antiparticle systems: the two energy terms with opposite sign $\pm E$ correspond to the two poles of the $S$ matrix and describe two completely different Majorana particle states with well-defined Majorana masses, opposite chirality states and with the conditions that each particle has to coincide with its own antiparticle, representing the solution of the Majorana equation of motion in Eq.~(\ref{majarray}).

Our work consists in completing and validating, with the use of the Majorana Tower, the results of the works in Ref \cite{sierra,castro}, finding a connection between the energy eigenstates of Majorana quanta from \cite{sierra} then expressed with second-order Bessel functions $K_\nu (x)$ that, through the Kontorovitch-Lebedev integral transforms and from the Mellin-Barnes integral representation of $K_\nu (x)$ in terms of the Riemann's zeta function. When the energy equation satisfy an identity, then one finds the zeros of zeta.
In the second part we use the Majorana equation in the $S$-matrix approach.
If the assumption in Ref. \cite{castro} is valid, all the zeros are on the critical line and then one can say that the use of the Majorana infinite-components equation could help a clever investigator like the Lieutenant Columbo to show the validity of the RH from the little things that had still to be tied up, that were already present in the literature, and finding ``who done it'' to formulate the best counting of primes, namely, Bernhard Riemann, this from an ideal collective contribution from Sierra, Castro-Perelman and then us to trace a route to the demonstration of the Riemann Hypothesis.

We then analyzed the dynamics of the poles of the $S$-matrix in a hyperbolic spacetime finding that the Hamiltonian containing the poles of $A(s,t,u)$ belongs to the class of $xp$ Hamiltonians as well and depend on the angular momentum $j$ of the Majorana quanta finding a relationship where $j$ must be real to satisfy a relationship recalling the route to the RH from the Hilbert-P\'olya's side.
We notice that the poles and zeros coincide in the pole dynamics when the angular momentum of the particle $j$ is real and the Hamiltonian is there Hermitian, confirming the results obtained with the first approach finding interesting correspondences to the requirements dictated by the Regge trajectories in a hyperbolic support, viz., that the angular momentum operator $j$ for a Hermitian operator satisfying the requirements to satisfy the RH must be real-valued.
Another interesting approach to the RH is revisited within the framework of the
special properties of $\theta$ functions, and CT invariance \cite{perelman1}, but for the moment it goes beyond the purpose of the present work.

In our opinion, the Majorana Tower is the most general grid that can be provided starting from the Poincar\'e group of transformation, let's say axiomatically, on the physical world. It is not a coincidence that this solution comes out in years during which the quantum-relativistic scheme is outlined in its fullness and before of the formulation of the perturbative ``ghost''. In short, it is a sort of super-constraint.

We argue this can close the circle to a sort of collective proof to the RH with the already existing contributions that can be found in Ref. \cite{sierra,castro}. It is fascinating that a mathematical conjecture from the mid-1800s, as expressed in Eq.~(\ref{counting}) finds its solution through a formal theory of theoretical physics like in the Hilbert-P\'olya approach and $S$-matrix case in ``the time of quanta'', a time frame similar to that seen in the case of Fermat's theorem. This correspondence between the ``atoms of mathematics'' and those of physics (like Majorana's quanta) brings us back to Kronecker's words and leads us to think that integer numbers are part of the world just like atoms, stones and stars: God made the integer numbers, everything else is human work (Die ganzen Zahlen hat der liebe Gott gemacht, alles andere ist Menschenwerk.).
We end with some questions that may give inspiration to the reader for future works. Does Rindler spacetimes points at the end to a de Sitter space? Does the Tower have something else to say to the gauge theories or do they remain separate schemes, where the dynamics of the poles replace what is not seen or known from the tower?
We hope someone will reply these answers soon, in future works.

\section*{Acknowledgements}
We acknowledge for the useful comments, suggestions and discussions to improve the manuscript M. V. Berry, G. Sierra, D. C. Brody, P. Betzios, O. Papadoulaki, O. Gaddam, C. Castro Perelman, C. C. Stacco and R. Spigler.
\\
We are grateful to the anonymous referees for their useful suggestions; 
FT acknowledges Peter Weibel, ZKM and IPSOEA Barbarigo.

\section*{Appendix}

\subsection{A brief discussion on quantum PT-symmetric systems} \label{appA}

Non-Hermitian physics and Parity-Time (PT) symmetry represent at all effects an emerging field of research. Quantum Mechanics, in its classical approach, is mainly based on Hermitian Hamiltonians. A clear example is the quantity energy, as ``Conservation of energy is a fundamental concept that shapes our understanding of physical reality'' \cite{El-Ganainy}. 

We know that closed quantum mechanical systems require a description with real eigenvalues of the energy and this is usually described by an Hermitian Hamiltonian. 
In other situations, instead of considering the dynamical system ``\textit{in toto}'', one can find a satisfactory description of a physical system in a subspace of the space that is used to describe the whole system.
In this case, energy can be described as a quantity exchanged between this specific subsystem and its environment, as in an open quantum system. In certain cases the eigenvalues can be complex-valued. 

An example is the pioneering work by Gamow in 1928; he was one of the first who described with complex energy eigenvalues an open quantum systems describing the tunneling through a potential barrier modeling the alpha decay of the atomic nucleus; this was the first successful application of quantum mechanics to a problem in nuclear physics \cite{gamow,gamow2}. The so-called Gamow vectors are the eigenvectors of the Hamiltonian with complex eigenvalues whose imaginary part describe exponentially decaying (or growing) states \cite{bohm}.
Here, the complex energy eigenvalues are used to describe the rate of escape of the alpha particles from the decaying atomic nucleus. 

Another example is the class of Parity-Time (PT) symmetric non-Hermitian Hamiltonians that belong to the class of pseudo-Hermitian Hamiltonians. They can admit complex eigenvalues as well. Non-Hermitian and in particular PT-symmetric quantum systems, can have either complex conjugate or purely real eigenvalues.

These systems are invariant under PT transformations and their spectral properties are related through PT-symmetry. Noteworthy, PT-symmetry is neither a necessary nor a sufficient condition for a Hamiltonian to have a real spectrum.

For this reason, one can think that a general PT-symmetric system could not be the exact tool required by the Hilbert-P\'olya approach to demonstrate the Riemann hypothesis.
Generic PT-symmetric systems do not exclude that at least one non-trivial zero of $\zeta(z)$ lays outside of the critical line, i.e., ${\Re{\zeta(t)}}_{\forall t | \zeta (t) = 0} = 1/2$ as instead an Hermitian Hamiltonian is expected to do.

As already said, there are PT-symmetric systems that have purely real eigenvalues and one can span this sub-class of the PT-symmetric systems to describe the distribution of the zeta zeros. In the literature one can find certain particular examples of one-dimensional non-Hermitian Hamiltonians, based on a variant of $H=xp$, which actually admit real spectra of eigenvalues \cite{bender} and whose eigenvalues can find the searched $1-1$ correspondence with the zeros of $\zeta (z)$.
The Hamiltonian, $\hat H_B$, there considered presents all eigenvalues that are real and correspond to the nontrivial zeros on the critical line. 
In this case, the Hamiltonian is pseudo-Hermitian \cite{mosta} with positive eigenvalues and PT-symmetric.

The uniqueness of the correspondence between the nontrivial zeros of $\zeta(Z)$ and the eigenvalues of the operator $\hat H_B$ is proved by the fact that the similarity transformation to a Hermitian counterpart and its inverse are both strictly bounded \cite{brody,brody2}.
This means that the operator is self-adjoint without any isolated exceptions. 

One of the problems for a quantum formulation of the RH is that the eigenstates of the PT-quantum Hamiltonian there discussed is not normalizable. 
The normalization cannot be a problem any more if the operator is defined with a Rigged Hilbert Space (RHS) formalism to draw the spectrum of the operator.
RHS is the mathematical language that can be found in the theory of scattering and decay like in first Gamow's studies and for the construction of spectral decompositions of chaotic maps \cite{antoine,delamadrid}. 

The Hamiltonian $\hat H_R$ belongs to the class of $xp$ Hamiltonians, where spatial coordinate and momentum are involved and, defined in a Rigged Hilbert space formalism, as in standard quantum mechanics, the eigenstates of the variables position and momentum present the same properties.
In this view, with the RHS method, all the eigenstates associated with the trivial zeros are eliminated. This because they do not belong to that Hilbert space and cannot have any influence on the eigenvalues of the $\hat H_B$ operator in that Hilbert space and ``for precisely the same reason it also expels all those eigenstates corresponding to the nontrivial zeros off the critical line (if they exist); and these will not affect the eigenvalues either'' \cite{brody3}. 

By resizing the Hilbert space with the RHS one then cannot have the possibility to make any inference about the possible nontrivial zeros off the critical line. 
This applies not only to the approach used with the PT-symmetric $\hat H_R$ with real eigenvalues, but also to many other cases discussed in literature. This finding actually set real doubts on the feasibility of the Hilbert-P\'olya approach.

Non-hermitian Hamiltonians with P-T symmetry and positive eigenvalues can find a corresponding hermitian adjoint operator describing the same dynamics with the same eigenvalues if and only if the whole set of eigenvalues is real and positive defined.
For any PT-symmetric Hamiltonian $\hat H$ satisfying this case, there could  exist an operator $\hat V$ that relates $\hat H$ to its Hermitian adjoint according to $VHV^{-1} = H^\dag$ \cite{Mannheim}.

\subsection{The Majorana Tower} \label{appB}.

The solution described by Majorana in 1932 in the attempt of getting rid of Dirac's negative energy solutions, also known as ``Majorana Tower'' \cite{Majorana:NC:1932}, was promising to bypass some problems of negative energy states initially found in the Dirac equation.

To describe the infinite-components equation by Majorana let us start from the original Dirac equation \cite{dirac}, 
\begin{equation}
i \hbar \frac{\partial \psi}{\partial t} = \left( c~\hat {\bm \alpha} \cdot  \hat {\bm p} + m c^2 \hat \beta \right) \psi
\label{diraceq}
\end{equation}
for which the wave-plane solutions are 
\begin{equation}
\psi_{p \pm} = K \left(\begin{array}{c}U \\ \frac{(c \hat {\bm \alpha} \cdot  \hat {\bm p} )}{mc^2 \pm E}U \end{array}\right) \frac{exp[ i [ (\hat {\bm p} \cdot \hat {\bm x} \mp E t) / \hbar]}{\sqrt{2 \pi \hbar^3}}
\end{equation}
and are labeled by the suffix $\pm$ because they correspond to positive and negative energy solutions $E = \pm \sqrt{m^2c^4+p^2c^2}$, viz., for each quantum state possessing a positive energy $E$, there is always a corresponding state with negative energy $- E$. The $2 \times 1$ column vector $U$ is a constant vector and $K \in \mathbb{R}$ a normalization constant. This is the already discussed sea of negative states of energy that finally found a correct interpretation in terms of the existence of antimatter, confirmed with the discovery of the positron, the anti-electron, by Anderson in 1932 \cite{anderson}.

Ettore Majorana generalized the Dirac equation to any spin value, giving a set of solutions with positive-valued masses by using the generators of the Poincar\'e - Lorentz group and imposing the operator $m c^2 \hat \beta$ to have a strictly positive spectrum through a non-unitary transformation. 
This was proved possible if this term were unitary and infinite dimensional, as it is in Majorana's solution.
In this view, any of these particles coincide with their corresponding antiparticles and will always be characterized by a spectrum of positive energy eigenvalues that depend on a denumerable spectrum of infinite (and discrete) spin values, showing a clear relationship between angular momentum and energy for each different particle at rest. 

The spin spectrum found obeys both the Bose-Einstein and the Fermi-Dirac statistics \cite{Majorana:NC:1932} showing a slightly different algebra from the Dirac and Clifford algebra of matrices in a Minkowski spacetime \cite{salin,bisiacchi69,esposito}.
The precise relationship between spin and energy of plane waves with positive masses due to the spin angular momentum/mass coupling in the Hamiltonian, in the laboratory frame, is then given by Eq.~(\ref{spinmass}).

Energy, mass and spin for a particle at rest are related by Eq.~(\ref{spinmass}); for the mass and the corresponding energies at rest one obtains a spectrum $E_r = mc^2 = \left(j + 1/2 \right) E_0$.
Conversely, the spin of each particle will depend on the energy and mass of the particle itself, 
\begin{equation}
j = \frac{mc^2}{E_0} - 1/2 .
\label{spineq}
\end{equation}
The solutions found so far correspond to plane waves derived from a relativistic transformation of the waves with null momentum in the laboratory reference frame. 


Majorana studied the orthogonal transformations of the Lorentz group of the variables $ct, x_i, y_i, z_i$ with determinant equal the unity, excluding the case of the determinant equivalent to $-1$.
These transformations are ruled by four Hermitian matrices for which $\gamma_\mu \gamma_\nu + \gamma_\nu \gamma_\mu = 2 \delta_{\mu \nu}$; $(\mu, \nu = 1, 2, 3, 4)$, even if it holds for general $n-$dimensional spaces.

This implies that all the solutions related via an arbitrary transformation can be considered as a fundamental solution: if the matrix $\alpha_i$ is a solution for that operator, then $\alpha _i' = S\alpha_i S^{-1}$ is a solution as well if $S$ is an arbitrary non-singular operator $(|S| \neq 0)$.
For the representation of the $\alpha_i$'s one has to use non-normal coordinate systems and replaces the orthonormality condition -- related to Hermiticity -- with that of linear independence of the vector basis (fundamental vectors). This is found in the original paper of 1932. Passing through a non-unitary transformation, that preserves in any case the Hermiticity, as shown in the original work by Majorana, one finds that the Hamiltonian is Hermitian.

The result is that the exactly same Hermitian matrices that have been obtained to describe Hermitian operators, with these non-orthogonal transformations they instead represent Hermitian operators. 
If one uses normal coordinate systems, the matrices that represent the searched Hermitian operators can be obtained from the set of fundamental matrices through a unitary (and the non-unitary $\widetilde{\varphi}\psi\varphi$) transformation.
The matrices representing non-Hermitian operators, instead, can be obtained from the same fundamental matrices through non-unitary transformations and in general these matrices will be non-Hermitian.

The Tower represents the simplest infinite-dimensional unitary representation of the Lorentz group, which is an open group and admits irreducible representations in infinite dimensions.
The fermionic and bosonic solutions to the Tower represent two classes of unitary representations of the Lorentz group. The spin is the representation index (of the group) and corresponds to the angular momentum number.

Any representation (fermionic or bosonic) acts on an infinite-dimensional space and the unitary vectors of the representation have two numbers $j=s$ and $m=j, j-1, ..., -j$, where $s$ can be the spectrum of fermions or bosons. A real number $Z_0$ represents the value of the group invariant that acts on the given representation. Any representation can be built from the real Lorentz transformations on the $4-$dimensional variables $(ct, x, y, z)$. These transformations can be deduced from the infinitesimal transformations associated to the matrices $a_x$, $b_x$ and $a_y$, $b_y$ and $a_z$, $b_z$ (S and T in Majorana 1932). 
The matrices $a$ and $b$ follow their commutation rules.
These matrices are Hermitian for unitary representations and non-Hermitian for non-unitary representations and $Z_0=\sum_{i=1}^3a_i b_i$.

Semiclassical analogies and physical interpretations of this mathematical structure can be found starting from the geometrically-``squared'' structure (in the Heisenberg-like geometric form) of the phase space assumed in the Berry-Keating model that can be interpreted as the set of values of the spin angular momentum of the particles found in the Majorana Tower.
The sum of the eigenfunctions of the $\hat H = xp + px \sim \sqrt{x} \hat p \sqrt{x}$.

To obtain the distribution of the values in the Majorana Tower according to the non-trivial zeros of $\zeta(z)$, one can consider the Rindler Hamiltonian in a manifold $D = 1 + 1$ is Hermitian for a Dirac spinor. 
One has to move to Majorana particle and relate z with the spin as $xp=s$. In this case the meaning of the spin in one dimension, as there are no rotations in the $D = 1 + 1$ space, only boosts are possible and correspond to the momenta and spin values found in the Tower in a Rindler spacetime. As operators we have $P^2$, which is the momentum squared and is the only Casimir operator of the Poincar\'e group in $d=1+1$. For the angular momentum the Casimir operator is $L^2$ with eigenvalue $L(L+1)$, this in general is valid for Weyl or Majorana spinors, but not for both in general, while in the $D = 1 + 1$ it is valid in both cases.
One way to build up a spin from the semiclassical operator $xp$ in $d=1+1$ dimensions, is to consider the gamma matrices that can be split into time plus the Pauli spin matrices; then project down from $3$ dimensions in the Pauli spin matrices down to $1$ dimension, namely to restrict the choice to the matrix $\sigma_z$ and leaving the other $\sigma$'s matrices off. The spin still exists in $D = 1 + 1$ as eigenvectors to $\sigma_z$ for the metric $ds^2= -z^2 dt^2 + dz^2$.

Following Majorana's original work, the Hamiltonian of the Tower is Hermitian even if it needs the non-unitary transformation to make the positive-energy/mass and negative energy/mass values coincide for both the bosonic and fermionic families of particles and thus this can satisfy the requirements of the HP approach. 

It is well-known that the general solutions of the time-dependent Dirac equation in $D = 1 + 1$ dimensions that may also include a Lorentz scalar potential and obey the Majorana condition -- in the Majorana representation -- are described by Hermitian Hamiltonians in the equation of motion \cite{devincenzo}. This means that in a $D = 1 + 1$ spacetime a Majorana particle at rest inside a box, or a free Majorana particle (or, better, in a penetrable box with periodic boundary conditions), including the case where the particle is inside an impenetrable box with no potential, and in a linear potential, are all described by an equation of motion $i \hbar \widehat{1}_2 \partial_t \Psi = \hat h \Psi$ where the Hamiltonian $\hat h$ is Hermitian and the energy eigenvalues are real-valued; $\widehat{1}_2$ is the $2 \times 2$ identity matrix.

In general, the Tower is always CPT invariant when one has a finite string of the angular momentum values $s$, unless finding a particular Hamiltonian where the lacking of negative energies is not a problem as in Ref. \cite{casalbuoni}. 
The same distribution can be found in a simpler way. The physical relevance of tachyonic-resonances/tachyonic-condensates that are found in the Majorana Tower correspond and are found also in bosonic string theory \cite{string}, where is found an important connection between string theory and the Riemann Hypothesis and Rindler spacetimes.

The physical interpretation of the Majorana solution is a wave vector that represents simultaneously bosons and/or fermions from their intrinsic momentum. If one changes reference frame with a Lorentz boost or a general relativistic transformation all the spinor components will result mixed together and cannot be easily separated unless adopting the lab frame, associated with the center of mass, where the spinor transforms like that of a particle with a defined spin. 

Concluding, the free Majorana field is CPT invariant and the solutions in the Rindler spacetimes for non-tachyonic particles are real and the spectrum of energy for the Hamiltonian is described by the Hamiltonian operator as in Eq.~(\ref{hsierra}).

\subsection{Hardy's Z and Wick rotations}
We can set a correspondence between the trivial and non trivial zeros of the Riemann $\zeta(z)$ and those of the Hardy $Z(t)$ through the Riemann-Siegel formula $\theta(t)$, interpreted in terms of a function of a set of rotations plus Lorentz boosts that connects, at a given specific event in the spacetime, the laboratory reference frame with a family of Rindler observers or transformations.

 The trivial zeros of $Z(t)$ are then mapped into the non-trivial zeros of $\zeta(z)$ through a transformation that corresponds to a $U(1)$ rotation in the complex plane of the Majorana solution. All the trivial zeros of $Z$ are real and correspond to the classical Majorana solution, an Hermitian Hamiltonian plus the Lorentz boost for the Rindler observer described by the Riemann-Siegel function $\theta(t)$ that preserves Hermiticity, in agreement with the prescriptions of Hilbert-P\'olya approach as also discussed in \cite{sierra}.

It is easy to show that this infinite spectrum of spin angular momenta can find a 1 to 1 correspondence with the set of trivial zeros of Riemann's $\zeta(z)$ through a simple bijective transformation $\{s_F\} \rightarrow \{\frac 1m s^{-1}\}$ and/or $\{s_B\} \rightarrow \{\frac 1{2m} (2s-1)^{-1}\}$ for the fermionic and bosonic subsets (that are also isomorphic to themselves), respectively.
Following this way, the standard Majorana tower can be considered isomorphic to the set of the trivial zeros of the Riemann $\zeta(z)$ function and to the  set of natural numbers, $\mathbb{N}$, with the form factor that depends on both the spin and mass correspondence as in Eq.~(\ref{spineq}). In fact, to characterize an infinite components wave function like the Majorana ``Tower'' one tries to get the mass spectrum and the form of factors. The existence and the structure of the form factors, the dynamic properties of the system can be seen as characteristic manifestations of the composite nature of the systems considered so far, like an hydrogen atom \cite{bisiacchi69}. 

If one deals instead with real eigenvalues given by Hermitian Hamiltonians, one can find a further correspondence between the real solutions and complex-Hermitian operators describing the zeros of $\zeta(z)$ that can be obtained with a phase operator acting on the variable $t$, corresponding to a spacetime rotation and acts as a Wick rotation between the scattering poles and the nontrivial zeros of $\zeta(z)$.
The $\zeta(z)$ standard zeros mapped by the standard Majorana solution,
\begin{equation}
Z(t)=e^{i \theta(t)}\zeta \left(\frac 12 + i t \right)
\end{equation}
where $\theta(t)$ is the Riemann-Siegel function (or Hardy's function), also known as the phase of the zeta function \cite{sierra}, here acting as a phase factor function \cite{edwards} that can be expressed also in terms of a  Veneziano amplitude \cite{he15}
\begin{equation}
e^{i \theta(t)}= \pi^{-it/2}\frac{\Gamma(1/4+it/2)}{|\Gamma(1/4+it/2)|}.
\end{equation} 
Hardy's $Z$ is an even function, real analytic for real values and maps the trivial Riemann zeros into his non-trivial zeros and vice-versa, with the interesting property that $\forall t \in \mathbb{R} \Rightarrow Z(t) \in \mathbb{R}$. 
The Riemann-Siegel function $\theta (t)$ acts as an operator, a phase-like rotator in the $U(1)$ symmetry, which is locally equivalent to a rotation in $\mathbb{R}^2$ of the special orthogonal group $SO(2)$ of real numbers and can be interpreted as a function of Lorentz boosts in a $D = 1 + 1$ spacetime.
$U(1)$ is the set of complex numbers, or $1\times1$ complex matrices, such that $ z \bar z = \bar z z~=~1$. The unitary complex group preserves Hermiticity. 

The Riemann-Siegel $\theta(t)$ and the Riemann $\zeta (t)$ are both holomorphic in the critical strip.
Instead of finding the zeros of $\zeta(z)$ on the critical line one finds the sign changes of $Z$. The Riemann hypothesis is false if $Z$ has a negative local maximum or a positive local minimum.
This because of the ``rotation'' factor $e^{i \theta (t)}$ given by the Riemann-Siegel formula.
$\theta(t)$ is a holomorphic function of t, as shown by
\begin{equation}
\theta(t) = - \frac i2 \left[\ln \Gamma \left( \frac 14 + i \frac t2 \right) - \ln \Gamma \left( \frac 14 - i \frac t2 \right)\right] -\frac {ln(\pi)t}2
\end{equation}
noteworthy the $Z$ function on the critical line is real when the term $i \sin(\theta(t))=0$, giving as output the set of Gram points where $\theta(t)/\pi \in\mathbb{Z}$ that cannot be used to univocally determine the exact distribution of the $\zeta(z)$ zeros as not all the zeros are separated by Gram points \cite{gram,hutch}.

Majorana studied the orthogonal transformations of the Lorentz group of the variables $ct, x_i, y_i, z_i$ with determinant equal the unity, excluding the case of the determinant equivalent to $-1$.
These transformations are ruled by four Hermitian matrices for which $\gamma_\mu \gamma_\nu + \gamma_\nu \gamma_\mu = 2 \delta_{\mu \nu}$; $(\mu, \nu = 1, 2, 3, 4)$, even if it holds for general $n-$dimensional spaces.

This implies that all the solutions related via an arbitrary transformation can be considered as a fundamental solution: if the matrix $\alpha_i$ is a solution for that operator, then $\alpha _i' = S\alpha_i S^{-1}$ is a solution as well if $S$ is an arbitrary non-singular operator $(|S| \neq 0)$.
For the representation of the $\alpha_i$'s one has to use non-normal coordinate systems and replaces the orthonormality condition -- related to Hermiticity -- with that of linear independence of the vector basis (fundamental vectors). This is found in the Majorana's original work of 1932. Passing through a non-unitary transformation, that preserves in any case the Hermiticity, as shown by Majorana \cite{Majorana:NC:1932}, one finds that the Hamiltonian is Hermitian.
The result is that the exactly same Hermitian matrices that have been obtained to describe Hermitian operators, with these non-orthogonal transformations they instead represent Hermitian operators. 
If one uses normal coordinate systems, the matrices that represent the searched Hermitian operators can be obtained from the set of fundamental matrices through a unitary (and the non-unitary $\widetilde{\varphi}\psi\varphi$) transformation.
The matrices representing non-Hermitian operators, instead, can be obtained from the same fundamental matrices through non-unitary transformations and in general these matrices will be non-Hermitian.

The tower represents the simplest infinite-dimensional unitary representation of the Lorentz group, which is an open group and admits irreducible representations in infinite dimensions.
The fermionic and bosonic solutions to the tower represent two classes of unitary representations of the Lorentz group. The spin is the representation index (of the group) and corresponds to the angular momentum number.

Every representation (fermionic or bosonic) acts on an infinite-dimensional space and the unitary vectors of the representation have two numbers $j=s$ and $m=j, j-1, ..., -j$, where $s$ can be the spectrum of fermions or bosons. A real number $Z_0$ represents the value of the group invariant that acts on the given representation. Any representation can be built from the real Lorentz transformations on the $4-$dimensional variables $(ct, x, y, z)$. These transformations can be deduced from the infinitesimal transformations associated to the matrices $a_x$, $b_x$ and $a_y$, $b_y$ and $a_z$, $b_z$ (S and T in Majorana 1932) \cite{Majorana:NC:1932}. 
The matrices $a$ and $b$ follow their commutation rules.
These matrices are Hermitian for unitary representations and non-Hermitian for non-unitary representations and $Z_0=\sum_{i=1}^3a_i b_i$.

\textbf{A physical interpretation:} Semiclassical analogies and physical interpretations of this mathematical structure can be found starting from the geometrically-``squared'' structure (in the Heisenberg-like geometric form) of the phase space assumed in the Berry-Keating model that can be interpreted as the set of values of the spin angular momentum of the particles found in the Majorana tower.
The sum of the eigenfunctions of the $\hat H = xp + px \sim \sqrt{x} \hat p \sqrt{x}$.

To obtain the distribution of the values in the Majorana tower according to the non-trivial zeros of $\zeta(z)$, one can consider the Rindler Hamiltonian in a manifold $D = 1 + 1$ is Hermitian for a Dirac spinor. 
One has to move to Majorana particle and relate z with the spin as $xp=s$. In this case the meaning of the spin in one dimension, as there are no rotations in the $D = 1 + 1$ space, only boosts are possible and correspond to the momenta and spin values found in the tower by a Rindler observer. As operators we have $P^2$, which is the momentum squared and is the only Casimir operator of the Poincar\'e group in $d=1+1$. For the angular momentum the Casimir operator is $L^2$ with eigenvalue $L(L+1)$, this in general is valid for Weyl or Majorana spinors, but not for both in general, while in the $D = 1 + 1$ it is valid in both cases.
One way to build up a spin from the semiclassical operator $xp$ in $d=1+1$ dimensions, is to consider the gamma matrices that can be split into time plus the Pauli spin matrices; then project down from $3$ dimensions in the Pauli spin matrices down to $1$ dimension, namely to restrict the choice to the matrix $\sigma_z$ and leaving the other $\sigma$'s matrices off. In this case spin still exists in $D = 1 + 1$ as eigenvectors to $\sigma_z$ for the metric $ds^2= -z^2 dt^2 + dz^2$.

Consider the Rindler Hamiltonian,
\begin{eqnarray}
\hat H_R &=&\left(\begin{array}{cc} -i \left(z \partial_z +\frac12 \right) & mz \\mz & i \left( z \partial_z + \frac12 \right) \end{array}\right) =
\\
&=& \sqrt{z} \hat p_z \sqrt{z} \sigma_z + m z \sigma_x
\end{eqnarray}
this represents energy and angular momentum as in the Majorana tower, namely an angular momentum state plus a boost,
\begin{equation}
 \sigma_x=\gamma^0=\left(\begin{array}{cc}0 & 1 \\1 & 0\end{array}\right), \, \, \, 
 -i \sigma_y=\gamma^1=\left(\begin{array}{cc}0 & -1 \\1 & 0\end{array}\right).
\end{equation}
\\
The Dirac equation is
\begin{equation}
(\gamma^\mu \partial_\mu + i \chi) \psi = 0
\end{equation}
and the matrices are $\gamma^0 \gamma^0 = \mathbf{1}$, $\gamma^0 \gamma^1 + \gamma^1 \gamma^0 = \mathbf{0}$, $\gamma^1 \gamma^1 = - \mathbf{1}$.
The matrix $\mathbf{G} = \gamma^0$ makes Hermitian the $\gamma$'s. There are several representations to move from the Dirac to the Majorana solutions in $d=1+1$.
The Rindler space is obtained with $x^0 = z~\sinh t$, $x^1 = z~\cosh t$, now $(z,t)$ are the Rindler $D =1+1$ set of coordinates including four-point bosonic string scattering and its the crossing symmetrized amplitude, Veneziano amplitude $A(s,t,u)$, the four-point amplitude with s, t, and u Mandelstam variables \cite{he15}.

This is equivalent to a rotation given by the Riemann-Siegel rotation operator $RS(t)=e^{i \theta (t)}$ -- expanded in Lorentz boosts -- from the trivial zeros of the Riemann $\zeta(z)$ mapped into the non-trivial zeros of $\zeta(z)$ via the Hardy function $Z(t)$. The operator transforms the trivial zeros of $\zeta(z)$ corresponding to the solutions of the original spectrum of the Majorana Tower into the zeros for $Z(t)$ that correspond to those seen by the Rindler observer. The operator $RS(t)$ behaves as a rotation plus a boost in a $D = 1 + 1$ spacetime.


\end{document}